\documentclass[a4paper, 11pt]{article}
\pdfoutput=1
\usepackage{jheppub}

\usepackage[T1]{fontenc}

\usepackage{csquotes}

\addtocontents{toc}{\protect\setcounter{tocdepth}{2}}

\title{Probing the Singularity of Scalar-Haired Black Holes with Holographic Complexity}
\author[a]{Giuseppe Policastro,}
\author[a, b, c]{Simon Wittum}
\affiliation[a]{Laboratoire de Physique de l'\'{E}cole Normale Supérieure, ENS, Universit\'{e} PSL, CNRS, Sorbonne Universit\'{e}, Universit\'{e} de Paris, F-75005 Paris, France}
\affiliation[b]{Departement Physik, ETH Zürich, CH-8093 Zürich, Switzerland}
\affiliation[c]{D\'{e}partement de Physique, École Polytechnique, IP Paris, F-91128 Palaiseau, France}
\emailAdd{giuseppe.policastro@phys.ens.fr}
\emailAdd{simonwittum@gmx.de}

\abstract{%
    It has been shown that the \enquote{complexity=anything} observables allow more possibilities to probe the geometry behind the horizon of
    AdS black holes compared to the volume complexity. For uncharged black holes, these observables access the geometry all the way to the vicinity of the singularity, while for charged black holes, they only probe
    up to the inner horizon. Under appropriate conditions, the near-singularity geometry takes the universal form of a Kasner spacetime, characterized by the
    Kasner exponents. By introducing scalar hair, it is possible to continuously vary the Kasner exponents away from their vacuum values. 
    In this work, we study the behavior of two different   observables to determine whether they remain viable holographic duals of complexity in the presence of 
    scalar hair. We also investigate how deeply these observables can probe the Kasner regime near the singularity. To this end, we consider two scalar potentials: an exponential 
    potential, which admits analytic solutions, and a pure mass term, which requires numerical analysis.
}

\begin{document}
\maketitle
\flushbottom

\tableofcontents
\section{Introduction}

In a series of recent papers, (quantum computational) complexity has gained significant attention in the study of black holes and quantum gravity in the context of AdS/CFT \cite{Susskind:2014rva, Brown:2015bva, 
Brown:2015lvg}. The notion of complexity comes from the realm of quantum information and roughly speaking provides a measure for the difficulty of performing a quantum task. More precisely, 
a quantum algorithm can be represented as a unitary operator $U$ mapping an input state to an output state. Such an operator can be realized as a sequence of elementary quantum gates 
(CNOT, Hadamard, phase, etc.) drawn from a universal set. The complexity $\mathcal{C}(U)$ is defined as the minimum number of gates
from the universal set required to construct $U$ up to a specified tolerance. By definition, $\mathcal{C}(U)$ depends both on the choice of universal gate set and the tolerance.
Additional refinements can be made by adding a different cost to different types of gates, which can be motivated by physical difficulties in building certain gates \cite{Chapman:2021jbh}. It is also possible to define the complexity of a quantum state, by minimizing the complexity of unitary operators that output the state from a reference state given as input.

In chaotic systems such as black holes \cite{Maldacena:2015waa}, the complexity of a typical state is expected to have a linear growth in time, for a time exponential in the number of degrees of freedom. This timescale vastly exceeds other characteristic timescales, such as the scrambling time
\cite{Sekino:2008he}, the thermalization time, or the saturation time of entanglement entropy \cite{Hartman:2013qma}.
The latter quantities, which have a well-established dual in the holographic dictionary, cannot capture the fine-grained information on the state associated to its complexity. On the other hand, in the spacetime of AdS black holes, there are bulk quantites, such as 
the volume of the Einstein--Rosen bridge or the action of the Wheeler--DeWitt patch, that exhibit the same growth behavior. This observation motivated the original proposals of \enquote{complexity=volume} (CV) \cite{Susskind:2014rva} and \enquote{complexity=action} (CA) \cite{Brown:2015bva, Brown:2015lvg}. 

The authors of \cite{Belin:2021bga, Belin:2022xmt} have extended these proposals by introducing a broad class of new bulk quantities, dubbed 
\enquote{complexity=anything} (denoted by CAny in the following\footnote{This short notation was suggested to us by R.C. Myers.})  observables, which are evaluated on codimension-zero or -one regions of spacetime and also exhibit the desired linear late-time growth. The question arises of how one can differentiate between these different observables. One conceivable scenario is that the variety of CAny observables is in correspondence with different complexity measures, distinguished e.g. by the choice of the penalty factors. However, we do not know yet how to test this conjecture. 

Lacking a more concrete strategy, we can nevertheless explore the space of observables and try to find distinguishing features when  they are applied to different situations. Along these lines, it is of particular interest to understand whether complexity can be used to probe the structure of the black hole singularity. We recall that in the CV proposal, the extremal surface that gives the late-time complexity remains at a finite distance from the singularity. It has been shown that some of the generalized observables, 
in the uncharged black hole case, can probe regions in the interior arbitrarily close to the singularity;  
in the charged case, however, the observables can probe only up to the inner horizon \cite{Jorstad:2023kmq}. 
The potential of the CAny observables to probe behind horizons has been further investigated in \cite{Jiang:2025qai} for Bardeen black holes with multiple horizons.

In the present paper we extend the analysis of \cite{Jorstad:2023kmq} in two directions. First, we consider two different codimension-one CAny observables, given by adding to the volume a term proportional to $C^2$ (the square of the Weyl tensor) or $K$ (the trace of the extrinsic curvature), both scaled by a coupling constant. These are representatives of two classes of functionals: those that depend only on the pullback of geometric bulk scalars to the surface, and those that depend explicitly on the embedding of the surface in the bulk. 

Second, we consider the behavior of both of these functionals in black holes with scalar hair. In the vicinity of the singularity, the metric of these black holes
can be described by a Kasner form, characterized by the Kasner exponents. It has been shown that the presence of scalar hair allows for a continuous shift of the Kasner exponents from their
vacuum values \cite{Frenkel:2020ysx}. We study how deeply the observables can approach the singularity, and the correlation of their behavior with the Kasner exponents. We consider two different models for scalar-haired black holes: in one case the scalar potential is an exponential $\propto e^{\alpha \phi}$; the corresponding (Chamblin--Reall) black hole solutions are known analytically, but they are not asymptotically AdS \cite{Chamblin:1999ya}. In the other case, the scalar potential is a simple mass term; the solutions have AdS asymptotics, but they can only be found numerically. 

The late-time behaviors of the $C^2$- and $K$-observable have already been studied in hairless black holes. The main results that we obtain by adding scalar hair are the following:
\begin{enumerate}
    \item The $C^2$-observable exhibits linear late-time growth only within a narrow window of the coupling constant for hairless black holes. We find numerically that this window widens in the Chamblin--Reall background but shrinks in the massive scalar background, eventually leaving only the volume functional with the expected late-time behavior (see Figures~\ref{fig:c2_scan} and \ref{fig:massive_slices}).
    
    \item The $K$-observable maintains linear late-time growth for all coupling values in hairless black holes. Our studies show that this property persists in both the Chamblin--Reall and massive scalar backgrounds.
    
    \item For sufficiently large negative couplings, the $K$-observable can probe the future singularity in hairless black holes. In the scalar-haired backgrounds, this effect is enhanced: extremal late-time surfaces can approach the singularity more closely as the strength of the scalar hair increases (see Figures \ref{fig:cr_slices} and \ref{fig:massive_slices}, bottom-left panels).
    
    \item While the late-time growth rate of the $K$-observable is symmetric under sign reversal of the coupling constant in hairless black holes, this symmetry is broken in the scalar-haired backgrounds. Negative couplings, which correspond to deeper interior probes, produce larger growth rates, with the effect most pronounced when the Kasner exponents deviate maximally from their vacuum values (see Figures \ref{fig:cr_slices} and \ref{fig:massive_slices}, bottom-right panels, and \ref{fig:kasner_probes}).
\end{enumerate}

The structure of this paper is as follows: In Section~\ref{sec:model}, we introduce the holographic model of a scalar-haired black hole and study it for the two different potentials.
Section~\ref{sec:complexity} reviews the codimension-one CAny observables proposed in \cite{Belin:2021bga, Belin:2022xmt}. In particular, we study the two observables mentioned above and discuss their behavior in the limit of vanishing scalar hair.
Section~\ref{sec:probes} extends the analysis to the case where hair is added. In particular, we study how far the observables can probe into the black hole, and how this affects their late-time growth rates. Finally, we summarize our findings and discuss future directions in Section~\ref{sec:conclusion}. In addition, we provide three appendices. Appendix~\ref{sec:a_boundary_contributions} computes the boundary contributions to the $C^2$- and $K$-observables in the Chamblin--Reall background, which differ from those in asymptotically AdS spacetimes. In Appendix~\ref{sec:b_late_time}, we analyze the time evolution of the $C^2$-observable and discuss the existence of extremal surfaces at late times. This is repeated in Appendix~\ref{sec:c_time_evolution_k} for the $K$-observable, where we additionally provide a geometric interpretation of the evolution of the extremal surfaces, which explains why they are able to probe the singularity. Both Appendix~\ref{sec:b_late_time} and Appendix~\ref{sec:c_time_evolution_k} focus on hairless backgrounds for simplicity, as no qualitative changes are expected to occur when hair is added.
\section{Black Holes with Scalar Hair}
\label{sec:model}

We consider a scalar field $\phi$ in $d+1$ spacetime dimensions minimally coupled to Einstein--Hilbert gravity and subject to the potential $V(\phi)$. 
The theory is described by the action
\begin{equation}
    S = \int d^{d+1}x \sqrt{-g} \left(\frac{1}{2\kappa_{d+1}^2}R-\frac{1}{2}g^{\mu\nu}\nabla_\mu \phi \nabla_\nu \phi - V(\phi)\right),
    \label{eq:action}
\end{equation}
where $R$ is the Ricci scalar and $\kappa_{d+1}$ the gravitational coupling constant which we are setting to 1 in the following. A potential 
cosmological constant has been absorbed in the definition of $V(\phi)$. The equations of motion are given by
\begin{subequations}
    \label{eq:einstein_eqs}
    \begin{align}
        R_{\mu\nu} - \frac{1}{2} R g_{\mu\nu} &= T_{\mu\nu}
        = \nabla_\mu \phi \nabla_\nu \phi - g_{\mu\nu} \left( \frac{1}{2} g^{\alpha\beta} \nabla_\alpha \phi \nabla_\beta \phi + 
        V(\phi) \right), \label{eq:einstein1}\\
        \nabla^2 \phi - V'(\phi) &= 0.
        \label{eq:einstein2}
    \end{align}
\end{subequations}
We are interested in static black hole solutions with planar horizon topology. Therefore, we choose to work with the ansatz
\begin{equation}
    ds^2 = \frac{1}{z^2} \left( -\rho(z) e^{-\chi(z)} dt^2 + \frac{1}{\rho(z)} dz^2 + \sum_{i=1}^{d-1}dx_i^2 \right)
    \quad \text{and} \quad 
    \phi = \phi(z).
    \label{eq:ansatz}
\end{equation}
Substituting this ansatz back into the equations of motion leads to a coupled system of two ordinary differential equations for $\phi$ and
$\rho$, and a third independent equation for $\chi$, given by
\begin{subequations}\label{eq:eom}
\begin{align}
    \phi'' &= \frac{2(d-1)z \rho \phi' - 2z^2 \rho' \phi' + 2V' + \frac{2}{d-1}z^3\rho\phi'^3}{2 z^2 \rho}, \label{eq:eom1} \\
    \rho' &= \frac{d(d-1)\rho + 2V + z^2 \rho \phi'^2}{(d-1) z}, \label{eq:eom2} \\
    \chi' &= \frac{2}{d-1} z (\phi')^2, \label{eq:eom3}
\end{align}
\end{subequations}
where primes denote differentiation with respect to $z$. The integration constant for $\chi$ obtained by solving Equation~\eqref{eq:eom3} is not physically relevant as
it can be absorbed in the definition of the time coordinate, leaving us with three conditions yet to be fixed. Since black hole solutions are characterized by the presence of an
event horizon at $z=z_\text{h}$, we need to set $\rho(z_\text{h})=0$, which fixes one condition. It can be seen that
Equation~\eqref{eq:eom1} diverges at the horizon unless we set
\begin{equation}
    \phi'(z_\text{h}) = \frac{V'(\phi(z_\text{h}))}{z_\text{h}^2 \rho'(z_\text{h})},
\end{equation}
fixing another condition.
Regular black hole solutions to Equations~\eqref{eq:eom} are therefore uniquely fixed by the values of $\phi(z_\text{h})=\phi_\text{h}$ and $z_\text{h}$.
Solutions with the same $\phi_\text{h}$ but different $z_\text{h}$ are related by a scaling transformation. To see this, consider the functions 
$\{\phi(z), \rho(z), \chi(z)\}$ solving Equations~\eqref{eq:eom} for given values of $\phi_\text{h}$ and $z_\text{h}$. One can show that the rescaled functions
$\{\phi(\lambda z), \rho(\lambda z), \chi(\lambda z)\}$ also satisfy the equations, with the horizon located at $z_\text{h}/\lambda$ and the value of the scalar field at the horizon
unchanged.

The presence of an event horizon gives a finite temperature to each solution due to Hawking \cite{Hawking:1975vcx}. To compute it, we use the Killing vector 
$\xi=\partial_t$ which is timelike outside, lightlike on, and spacelike behind the horizon. The 
Hawking temperature is determined by the surface gravity $\kappa$ at the horizon as
\begin{equation}
    T = \frac{\kappa}{2\pi} =  \frac{1}{2\pi}\sqrt{-\frac{1}{2}\nabla_\mu \xi_\nu \nabla^\mu \xi^\nu} = \frac{|\rho'(z_\text{h})|}{4\pi}e^{-\chi(z_\text{h})/2}.
    \label{eq:hawking_temperature}
\end{equation}
Although this formula involves quantities evaluated at the horizon, the resulting temperature $T$ is the one measured by an observer infinitely far away from the black hole. If the spacetime is
asymptotically AdS, this temperature can be identified with the temperature of the dual field theory state. Additionally, the entropy of the black hole is given by the Bekenstein--Hawking 
formula
\begin{equation}
    S = 2\pi A_\text{h}
    \quad \text{with} \quad
    A_\text{h} = \frac{1}{z_\text{h}^{d-1}} \int d^{d-1}x,
    \label{eq:bekenstein_hawking_entropy}
\end{equation}
where $A_\text{h}$ is the infinite area of the planar event horizon. The finite entropy density $s$ is obtained by dividing $S$ by the integral over the spatial coordinates,
yielding $s=2\pi z_\text{h}^{1-d}$.

\subsection{Chamblin--Reall Solution}
In \cite{Chamblin:1999ya}, Chamblin and Reall discovered an analytic black hole solution with a non-trivial scalar field profile for the case of an exponential potential given by
\begin{equation}
    V(\phi) = -\frac{d(d-1)}{2L_V} e^{\alpha \phi},
\end{equation}
where $L_V > 0$ and $\alpha$ are constants. In the limit $\alpha=0$, this potential becomes a cosmological constant and $L_V$ coincides with the AdS radius. 
We will set $L_V=1$ in the following. In our coordinate system, the equations of motion for this potential are solved by
\begin{subequations}\label{eq:cr_solutions}
    \begin{align}
        \phi(z) &= \phi_\text{h} + \frac{\alpha}{2}(d-1) \log\left(u\right), \\
        \rho(z) &= \frac{e^{\alpha \phi_\text{h}}}{1-\eta/(2d)}u^{\eta}\left(1-u^{d\left(1-\eta/(2d)\right)}\right), \label{eq:rho} \\
        \chi(z) &= \eta\log(u),
    \end{align}
\end{subequations}
with $u = z/z_\text{h}$ and $\eta = \alpha^2 (d-1)/2$. As expected, solutions with the same $\phi_\text{h}$ but different $z_\text{h}$ are related by a scaling transformation.
For $\alpha=0$, the solution reduces to a hairless black hole with $\phi=\phi_\text{h}$, $\rho=1-u^d$, and $\chi=0$. 
\begin{figure}[!htbp]
    \centering
    \includegraphics[width=0.95\textwidth]{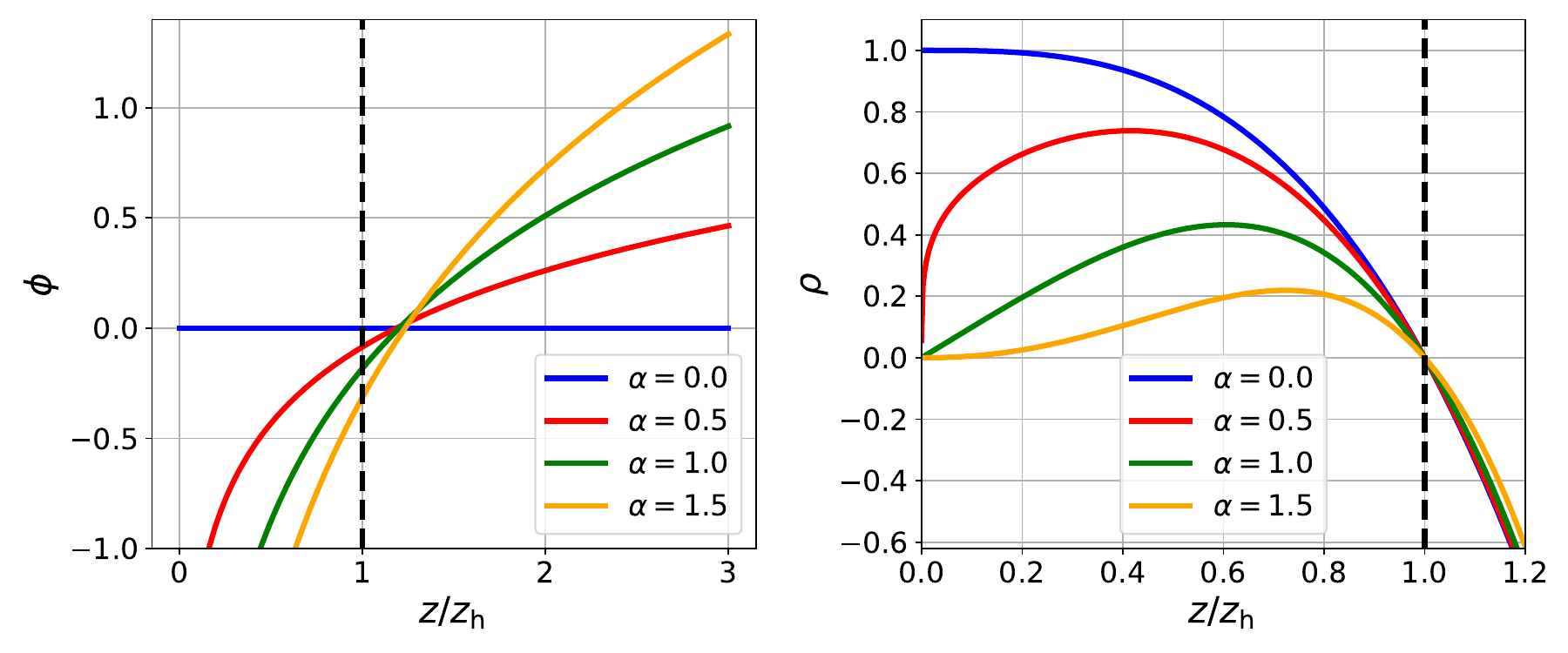}
    \caption{Chamblin--Reall solution for different values of $\alpha$ in $d=3$ spatial dimensions, with $\phi_\text{h}$ fixed according to Equation~\eqref{eq:phi_h}, showing
    $\phi(z)$ (left) and $\rho(z)$ (right). The vertical dashed line indicates the location of the horizon. Changing $\alpha$ deforms the Kasner exponents governing the near-singularity behavior, while Equation~\eqref{eq:phi_h} ensures that the 
    asymptotic form of the spacetime remains unchanged.}
    \label{fig:CR_solutions}
\end{figure}
Changing the value of the parameter $\alpha$ modifies the bulk geometry. However, unless $\phi_\text{h}$ is adjusted accordingly,
it also changes the metric on the conformal boundary, giving a relative rescaling of the time and space boundary coordinates. Since we want to compare different solutions within the same boundary class, $\phi_\text{h}$ needs to be chosen such that
the boundary metric remains invariant. This can be achieved by fixing the combination
\begin{equation}
    -z^2 g_{tt} = \rho e^{-\chi} \to \frac{e^{\alpha \phi_\text{h}}}{1-\eta/(2d)} = C,
\end{equation}
where $C$ is a constant characterizing the boundary class that is being considered. Solving this constraint for $\phi_\text{h}$, we obtain
\begin{equation}
    \phi_\text{h} = \frac{1}{\alpha}\left(\log(C)+\log\left(1-\eta/(2d)\right)\right).
    \label{eq:phi_h}
\end{equation}
We will work with $C=1$, which is the class that the hairless AdS black hole belongs to. With this choice, the transformation $\alpha \to -\alpha$ flips the sign of
$\phi_\text{h}$. The metric functions are invariant under this transformation and therefore, we can restrict to $\alpha > 0$.
The Hawking temperature of the spacetime can be computed using Equation~\eqref{eq:hawking_temperature}, yielding
\begin{equation}
    T = \frac{d e^{\alpha \phi_\text{h}}}{4\pi z_\text{h}}
    = \frac{d}{4\pi z_\text{h}} \left(1-\frac{\eta}{2d}\right),
\end{equation}
where we have used Equation~\eqref{eq:phi_h} in the second step.
In Figure~\ref{fig:CR_solutions}, the Chamblin--Reall solution is shown for different values of $\alpha$, 
with $\phi_\text{h}$ fixed according to Equation~\eqref{eq:phi_h}.
The plot demonstrates that $\rho(z)$ stays finite for $z\to 0$, while it diverges as $z\to \infty$. Therefore, the region $z\to 0$ is interpreted as the asymptotic region, while $z\to \infty$ 
corresponds to the singularity. For $\alpha\neq 0$, the solution does not asymptote to AdS. However, a holographic dictionary can still be established as the solution can be obtained by dimensional reduction 
of a higher dimensional system that does asymptote to AdS \cite{Gouteraux:2011qh}.

\subsubsection{Behavior Near the Singularity}
To study the behavior of the system near the singularity, we need to distinguish between two cases. For $\eta > 2d$, the solution can be brought into the form
\begin{equation}
        ds^2 = -dZ^2 + Z^{4/\eta}\left(dT^2+\sum_{i=1}^{d-1}dX_i^2\right)
        \quad \text{and} \quad
        \phi(Z) = -\frac{2}{\alpha} \log(Z) + \text{const.}
\end{equation}
by applying a suitable coordinate transformation. If $\eta < 2d$, the solution instead becomes
\begin{equation}
    ds^2 = -dZ^2 + Z^{2p_T}dT^2 + Z^{2p_X}\sum_{i=1}^{d-1} dX_i^2
    \quad \text{and} \quad
    \phi(Z) = -p_\phi \log(Z) + \text{const.},
\end{equation}
where the so called Kasner exponents have been defined as
\begin{equation}
    p_T = \frac{\alpha^2(d-1)-4d+8}{4d+\alpha^2(d-1)},
    \quad
    p_X = \frac{8}{4d+\alpha^2(d-1)}
    \quad \text{and} \quad
    p_\phi = \frac{4\alpha(d-1)}{4d+\alpha^2(d-1)},
\end{equation}
satisfying the two Kasner conditions
\begin{equation}
    p_T+(d-1)p_X = 1
    \quad \text{and} \quad
    p_\phi^2+p_T^2+(d-1)p_X^2 = 1.
    \label{eq:kasner_conditions}
\end{equation}
Only in this second regime the metric near the singularity becomes a genuine Kasner spacetime \cite{Kasner:1921zz, Belinskii:1973sud}, with $p_T=-(d-2)/d$, $p_X=2/d$ and $p_\phi=0$ for the hairless black hole.
As we are particularly interested in singularities that resemble a Kasner spacetime, we will restrict ourselves to this case for the remainder of the paper. 

For completeness, we note that in the critical case $\eta=2d$, the blackening function in Equation~\eqref{eq:rho} takes the shape
\begin{equation}
    \rho(z) = -de^{\alpha\phi_\text{h}} u^\eta \log(u),
\end{equation}
which does not simplify further close to the singularity. The logarithmic term prevents the metric from reducing to a power-law form, and no Kasner-like scaling regime emerges.

\begin{figure}[!htbp]
    \centering
    \includegraphics[width=0.95\textwidth]{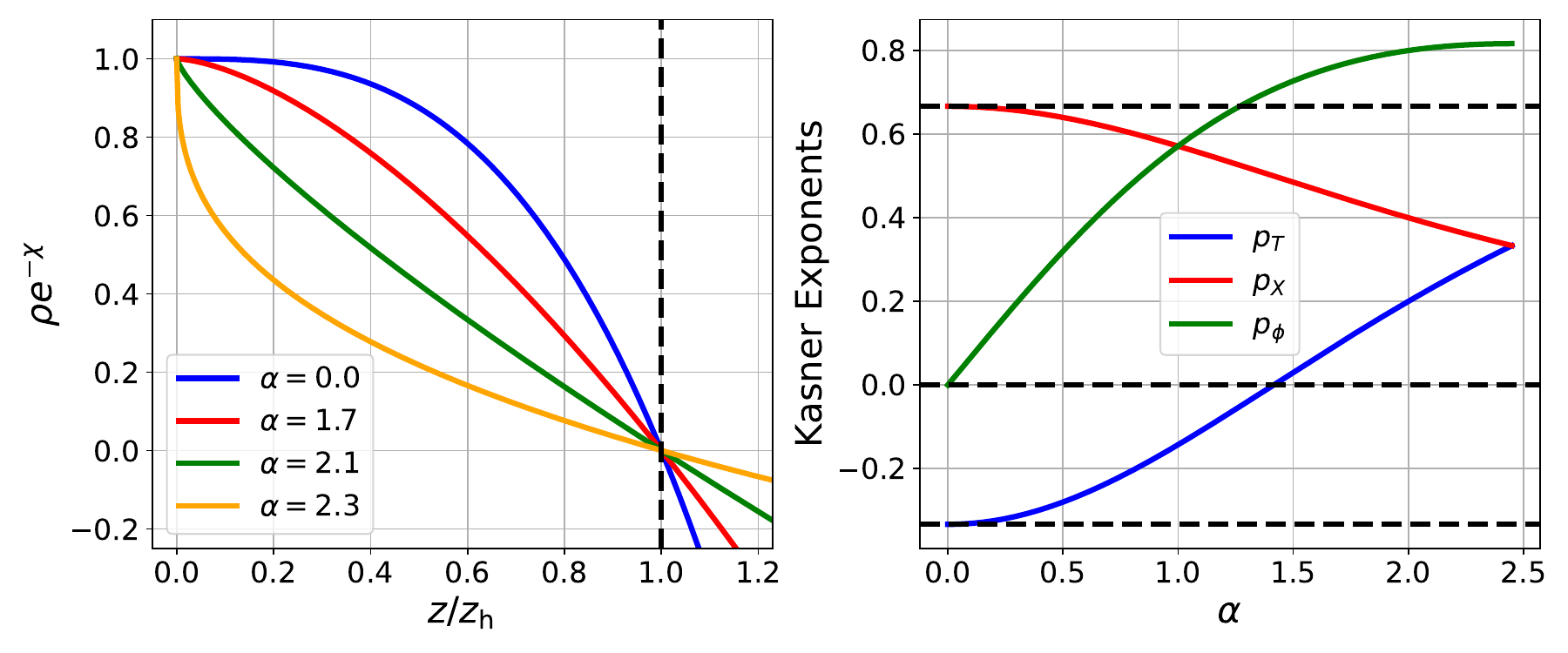}
    \caption{Left: Choosing $\phi_\text{h}$ as prescribed in Equation~\eqref{eq:phi_h} ensures that the asymptotic metric remains unchanged when $\alpha$ is modified.
    Right: The Kasner exponents, which characterize the near-singularity geometry and are independent of $\phi_\text{h}$, change monotonically with $\alpha$. 
    Both panels correspond to $d=3$ spatial dimensions.}
    \label{fig:kasner_cr}
\end{figure}

Since the Kasner exponents found for $\eta < 2d$ depend on $\alpha$ but not on $\phi_\text{h}$, tuning this parameter allows to modify the structure of the singularity without 
changing the asymptotic form of the metric, as shown in Figure~\ref{fig:kasner_cr}.

\subsection{Massive Scalar Solution}
The second potential we consider includes a mass term and a cosmological constant $\Lambda$ given by
\begin{equation}
    V(\phi) = \frac{1}{2} m^2 \phi^2 + \Lambda
    \quad \text{with} \quad
    \Lambda = -\frac{d(d-1)}{2L^2},
\end{equation}
where $L$ is the AdS radius that we will set to one. With a mass term of this form, no analytic solution is known that takes into account the backreaction of the scalar
field on the metric. Therefore, we need to resort to numerical methods and asymptotic expansions to study the behavior of the system. We will find that solutions to the equations of 
motion with this potential asymptote to AdS for $z\to 0$ and develop a Kasner singularity for $z\to \infty$.

\subsubsection{Behavior Near the Boundary}
We have seen that black hole solutions are characterized by the value of $\phi_\text{h}$. However, since $\phi_\text{h}$ is not directly
observable and therefore not particularly meaningful from the dual field theory perspective, we need to relate it to the boundary data, which we will
use to label the numerical solutions. In order to extract these parameters, an understanding of the boundary behavior has to be developed,
which is tricky for generic values of $d$ and $m^2$. Therefore, we will fix $d=3$ and $m^2=-2$ (above the Breitenlohner-Freedman bound \cite{Breitenlohner:1982jf}) for this part of the discussion
and the numerical analysis later on. With these choices, we have \cite{Frenkel:2020ysx}
\begin{equation}
    \phi(z) = J_{\mathcal{O}} z+\langle\mathcal{O}\rangle z^2+\dots
    \quad \text{and} \quad
    \rho(z) = 1+\frac{1}{2}J_\mathcal{O}^2z^2+\left(\frac{4}{3}J_\mathcal{O}\langle \mathcal{O}\rangle-\langle T_{tt} \rangle\right) z^3
    +\dots,
    \label{eq:asymptotic_boundary}
\end{equation}
where the dots indicate higher order terms in $z$. In the  of AdS/CFT, $\langle\mathcal{O}\rangle$ corresponds to the expectation value of the operator $\mathcal{O}$ defined on
the boundary CFT that is dual to the bulk scalar field $\phi$ and sourced by $J_\mathcal{O}$, and $\langle T_{tt} \rangle$ is the expectation value of the $tt$-component
of the stress-energy tensor in the dual field theory state. For the hairless AdS black hole, we have $\langle T_{tt} \rangle = z_\text{h}^{-3}$. Using Equation~\eqref{eq:eom3}, it is possible to obtain an
expansion for $\chi(z)$ as well. Combining the result with the asymptotic form of $\rho(z)$, we find
\begin{equation}
    -z^2 g_{tt}(z) = \rho(z)e^{-\chi(z)} = 1 - \langle T_{tt} \rangle z^3 + \dots,
\end{equation}
which will later be used to extract $\langle T_{tt} \rangle$ from the numerical solutions.

\subsubsection{Behavior Near the Singularity}
It can be shown that close to the singularity, the equations of motion are asymptotically solved by
\begin{equation}
    \phi(z) = C_\phi\log(z)+\dots, \quad
    \chi(z) = \frac{2C_\phi^2}{d-1}\log(z) + C_\chi + \dots, \quad
    \rho(z) = -C_\rho z^{d+\frac{C_\phi^2}{d-1}} + \dots,
    \label{eq:asymptotic_singularity}
\end{equation}
where $C_\phi$, $C_\chi$ and $C_\rho$ are integration constants. Substituting this back into the metric and performing a suitable coordinate transformation,
we find that the solution can be written as
\begin{equation}
    ds^2 = -dZ^2+Z^{2p_T}dT^2+Z^{2p_X} \sum_{i=1}^{d-1}dX_i^2
    \quad \text{and} \quad
    \phi(Z) = -p_\phi \log Z + \text{const.},
\end{equation}
where this time, the Kasner exponents are given by
\begin{equation}
    p_T = \frac{C_\phi^2-(d-1)(d-2)}{d(d-1)+C_\phi^2},
    \quad
    p_X = \frac{2(d-1)}{d(d-1)+C_\phi^2},
    \quad \text{and} \quad
    p_\phi = \frac{2(d-1)C_\phi}{d(d-1)+C_\phi^2},
    \label{eq:kasner_exponents}
\end{equation}
again satisfying the Kasner conditions given in Equation~\eqref{eq:kasner_conditions}. The value of $C_\phi$ depends on how the initial condition $\phi_\text{h}$ is chosen. Therefore,
tuning $\phi_\text{h}$ allows to modify the Kasner exponents and thus the structure of the singularity. This is different from the Chamblin--Reall solution, where the Kasner exponents
are fixed by the parameter $\alpha$ of the potential and do not depend on the initial conditions.

\subsubsection{Matching Boundary and Singularity Behavior}
\begin{figure}[!htbp]
    \centering
    \includegraphics[width=0.95\textwidth]{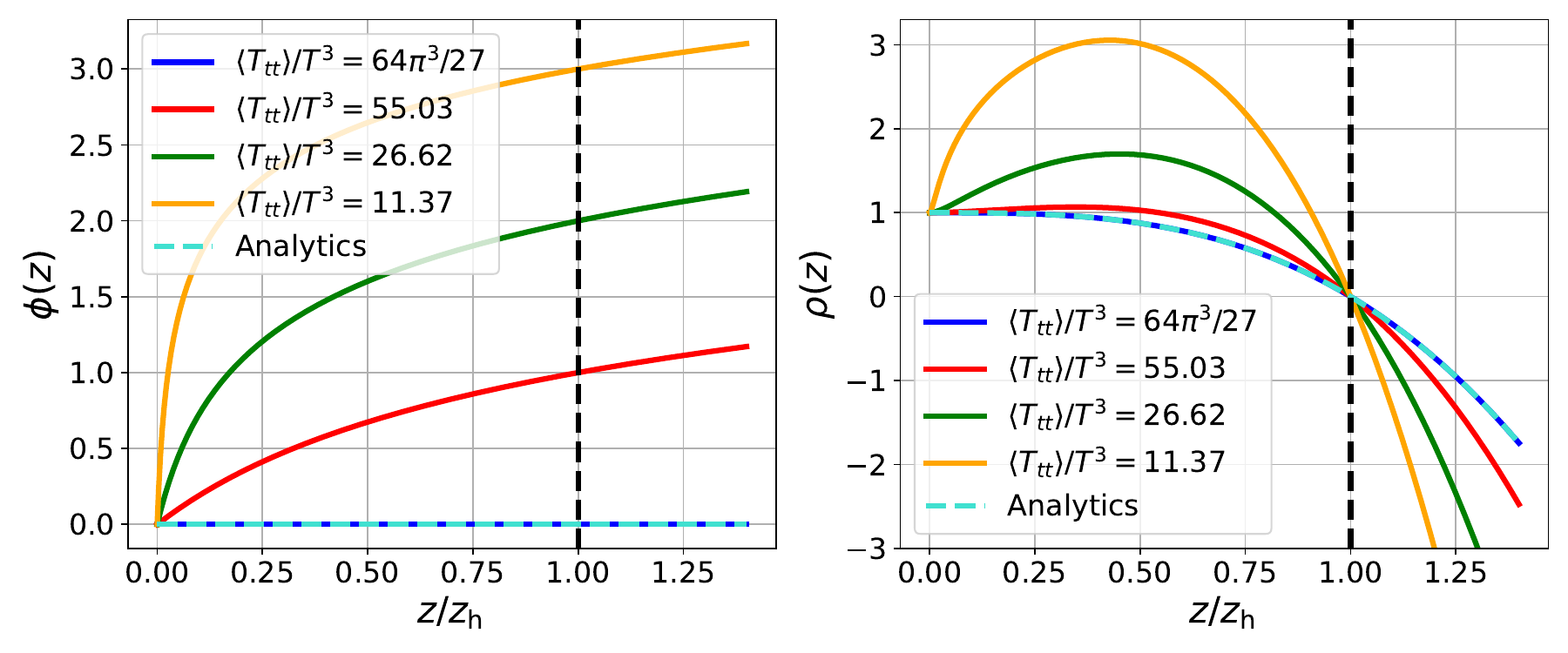}
    \caption{Numerical solutions to Equations~\eqref{eq:eom} for a massive scalar field with $d = 3$ and $m^2 = -2$, showing $\phi(z)$ (left) and $\rho(z)$ (right). 
    The solutions are obtained by integrating from the horizon toward both the boundary at $z \to 0$ and the singularity at $z \to \infty$ independently. Owing to regularity, the 
    values of $\phi(z)$ and $\rho(z)$ near the boundary are fixed, eliminating the need for a shooting method. As shown explicitly in Equation~\eqref{eq:asymptotic_boundary}, 
    both $\phi(z)$ and $\rho(z)$ diverge near the singularity. The vertical dashed line indicates the location of the horizon.}
    \label{fig:massive_scalar_solutions}
\end{figure}

With the asymptotic behavior known both near the boundary and close to the singularity, we can solve the equations of motion for various initial conditions to establish a connection
between boundary observables in the CFT and the Kasner exponents governing the near-singularity dynamics. As a demonstration, Figure~\ref{fig:massive_scalar_solutions} contains numerical
solutions for several values of $\langle T_{tt} \rangle/T^3$ for $d=3$ and $m^2=-2$. 
The scalar field is shown in the left panel. As expected from our discussion, it vanishes on the 
boundary and diverges logarithmically near the singularity. On the right, we see the metric function $\rho$, which approaches 1 at the boundary without the need to apply any shooting
method, and diverges polynomially close to the singularity. In both panels, the turquoise dashed lines represent the analytic hairless black hole solution with $\langle T_{tt} \rangle / T^3 = 64\pi^3/27$,
while the vertical dashed lines indicate the location of the horizon.

\begin{figure}[!htbp]
    \centering
    \includegraphics[width=0.95\textwidth]{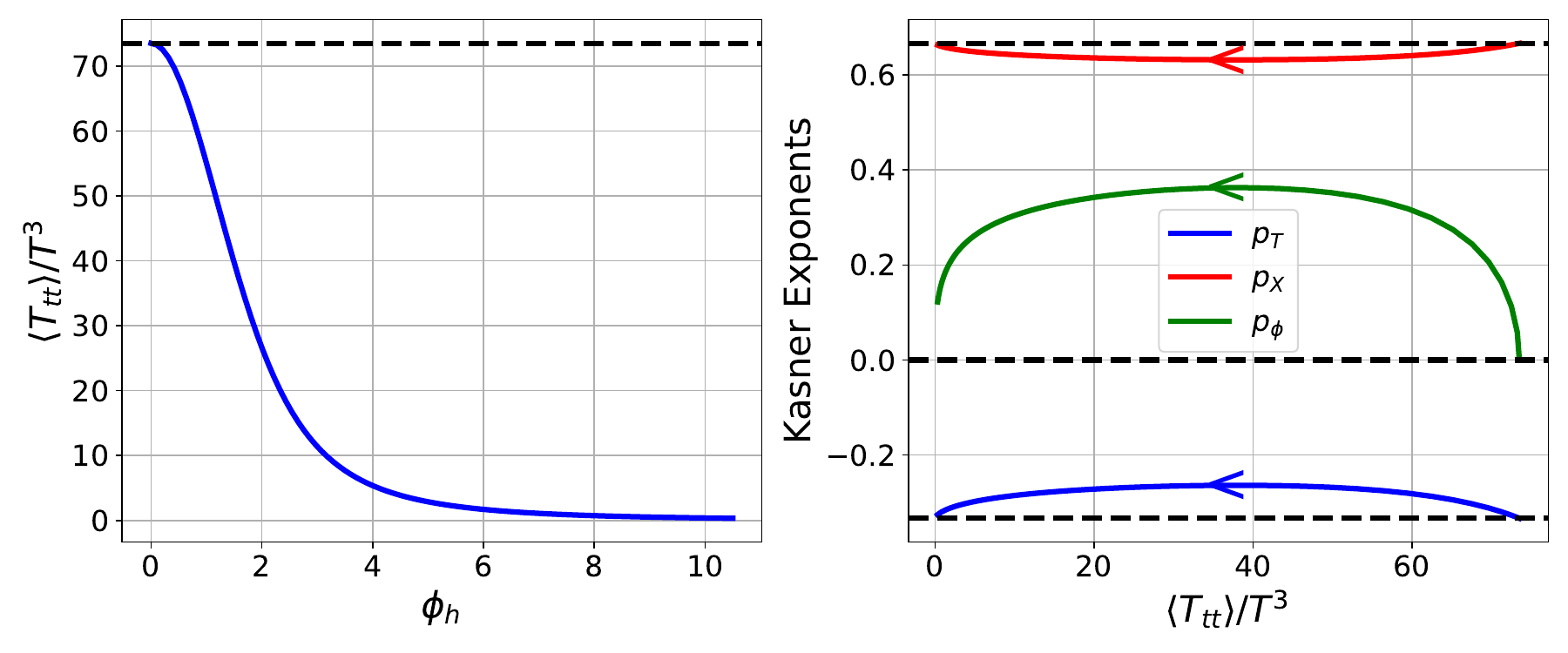}
    \caption{%
        Left: Plot of the normalized boundary energy density, 
        $\langle T_{tt} \rangle / T^3$, as a function of the scalar field's value at the horizon 
        $\phi_\text{h}$ in $d = 3$ dimensions for a pure mass potential with $m^2 = -2$. 
        Since the relation is monotonic, $\phi_\text{h}$ can be traded for the more physical label 
        $\langle T_{tt} \rangle / T^3$ to classify the numerical solutions. 
        The horizontal dashed line indicates the analytic result 
        $\langle T_{tt} \rangle / T^3 = 64\pi^3/27$ for $\phi_\text{h} = 0$.
        Right: Kasner exponents as a function of $\langle T_{tt} \rangle / T^3$.
        The dashed lines mark the reference values $p_X = 2/3$, $p_T = -1/3$, 
        and $p_\phi = 0$, corresponding to the solution without a scalar field, the arrows along the lines indicate the direction of flow as $\phi_\text{h}$ is increased.
    }
    \label{fig:Kasner_Exponents}
\end{figure}

By solving the equations of motion for different values of $\phi_\text{h}$, we obtain the monotonic relationship between $\phi_\text{h}$ and $\langle T_{tt} \rangle / T^3$
shown in the left panel of Figure~\ref{fig:Kasner_Exponents}. This monotonicity confirms that $\langle T_{tt} \rangle / T^3$ provides a suitable 
parameter to distinguish different solutions. We restrict the plot to positive values of $\phi_\text{h}$ because the equations of motion~\eqref{eq:eom} are symmetric under 
$\phi \to -\phi$ for a potential containing only even powers of $\phi$. We observe that increasing the field value at the horizon corresponds to a decrease in 
$\langle T_{tt} \rangle / T^3$.
For $\phi_\text{h}=0$, the numerics are able to reproduce the analytic result $\langle T_{tt} \rangle / T^3 = 64\pi^3/27$, as indicated by the dashed horizontal line.

Attempts to reconstruct the numerically obtained relationship between $\langle T_{tt} \rangle / T^3$ and $\phi_\text{h}$ in the low-temperature regime using the thermal gas 
phase (following the approach of \cite{Gursoy:2008za, Gursoy:2009jd}) have encountered inconsistencies.
These issues likely stem from the absence of a 
genuine low-temperature phase for the black hole solutions considered here;
in particular, we can confirm numerically that no solutions exist with $T<3/(4\pi z_\text{h})$, with the hairless solution satisfying $T=3/(4\pi z_\text{h})$.
Consequently, we cannot fully validate the numerical results beyond checking that they reproduce the free theory limit and remain positive, as expected for energy densities and temperatures.

The Kasner exponents are extracted by analyzing the behavior of the scalar field near the singularity, using Equations~\eqref{eq:asymptotic_singularity} and~\eqref{eq:kasner_exponents}.
Their dependence on $\langle T_{tt} \rangle / T^3$ is shown in the right panel of Figure~\ref{fig:Kasner_Exponents}. As expected, in the absence of a scalar field, the exponents take 
their vacuum values $p_X = 2/3$, $p_T = -1/3$, and $p_\phi = 0$, indicated by the horizontal dashed lines. When the scalar field is turned on, corresponding to a decrease of 
$\langle T_{tt} \rangle/T^3$, the exponents deviate from these vacuum values, reach extrema, and eventually return to the vacuum configuration.

\section{Holographic Complexity}
\label{sec:complexity}

\begin{figure}[!htbp]
    \centering
    \includegraphics[width=0.65\textwidth]{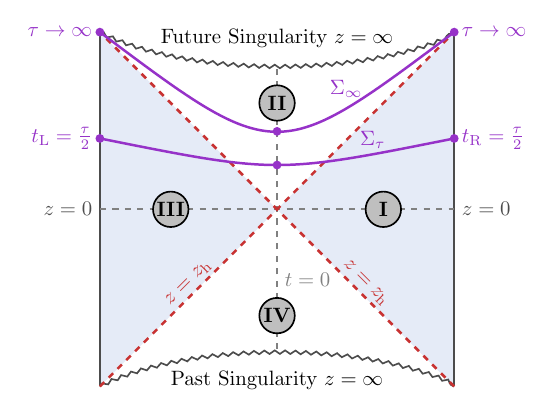}
    \caption{Penrose diagram of an eternal AdS black hole. For the planar black hole, each point represents a spatial slice of topology $\mathbb{R}^{d-1}$.
    The timelike boundaries of the exterior regions correspond to the domains of two independent CFTs. The extremal surfaces $\Sigma$ 
    connect the two boundaries and are conjectured to encode the complexity of the dual CFTs at the timeslice where they are anchored.}
    \label{fig:extremal_surface}
\end{figure}

We now turn to the study of holographic complexity in the background of the two black hole solutions introduced in the previous section. Consider a $d$-dimensional spacelike surface
$\Sigma$ parameterized by the embedding functions $X^\mu(\sigma^a)$ and anchored at a timeslice $\Sigma_\text{bnd}$ on the conformal boundary of spacetime, as visualized in
Figure~\ref{fig:extremal_surface}. The authors of \cite{Belin:2021bga} defined a family of observables $\mathcal{O}_\zeta[\Sigma_\text{bnd}]$ as the value of a 
functional $F_\zeta[\Sigma]$ evaluated on an extremal surface $\Sigma_\text{c}$, i.e., 
\begin{equation}
    \mathcal{O}_{\zeta}[\Sigma_\text{bnd}] = F_\zeta[\Sigma_\text{c}]
    \quad \text{with} \quad
    \delta \left( F_\zeta[\Sigma_\text{c}] \right) = 0,
    \quad \text{where} \quad
    F_\zeta[\Sigma] = \int_{\Sigma} d^d \sigma \sqrt{g} \zeta(X^\mu, g_{\mu\nu}).
\end{equation}
The dimensionless scalar function $\zeta(X^\mu, g_{\mu\nu})$ is a geometric invariant constructed locally from the ambient metric $g_{\mu\nu}$, 
the embedding $X^\mu(\sigma^a)$, and their respective derivatives. For notational simplicity, we omit the explicit dependence on the derivatives. In the case that multiple extremal surfaces
exist for the same boundary timeslice $\Sigma_\text{bnd}$, the prescription is to choose the surface that maximizes the value of the functional out of all extremal surfaces.

It was proposed that the observables $\mathcal{O}_\zeta$ can be interpreted as holographic duals of different complexity measures
of the boundary field theory on the timeslice $\Sigma_\text{bnd}$. The main motivation for this proposal is that for suitable choices of $\zeta$, the 
observables above exhibit the same linear late-time growth behavior as complexity, and that they exhibit the switchback effect in the presence of
shockwaves \cite{Jorstad:2023kmq}. Importantly, this proposal includes the CV proposal as a special case by choosing $\zeta(X^\mu, g_{\mu\nu}) = 1$. Additionally, the proposal
has been generalized to functionals evaluated on codimension-zero regions of spacetime, which includes the CA proposal as a special case \cite{Belin:2022xmt}, as well as observables defined by evaluating a functional $F$ on the surface which extremizes a different functional $G$. We will not consider these further extensions in this paper. 

In the following, we assume that the background metric is given by Equation~\eqref{eq:ansatz}. The Penrose diagram in figure~\ref{fig:extremal_surface} represents the causal structure of both classes of solutions discussed in Section~\ref{sec:model}.
Additionally, we fix $\Sigma_\text{bnd}$ to consist of two full planar slices at times $t_L$ and $t_R$ on the left and right boundaries, respectively, so the observable depends only on these times. However, 
due to the translation symmetry in $t$ generated by the Killing vector $\xi$, the value of the observable depends only on the sum $\tau = t_L + t_R$. Without 
loss of generality, we can therefore restrict to symmetric configurations with $t_L = t_R = \tau/2$, and denote the corresponding surface selected by the above procedure $\Sigma_\tau$. 
Because of this symmetry, it is sufficient to compute the variation of half of the 
surface, which is helpful since coordinates covering the full spacetime are generally not analytically tractable. The $z$-coordinate of the gluing point $z_\text{max}$ at $t=0$ can be 
fixed by computing all extremal half-surfaces as a function of the gluing point, and plugging the result back into the functional. The value of $z_\text{max}$ is then determined 
by finding the extremal point of this function. It can be shown that this is equivalent to the condition that the surface is smooth at the gluing point. In Figure~\ref{fig:extremal_surface}, 
the endpoints of the half-surface are indicated by the purple blobs.

The metric in Equation~\eqref{eq:ansatz} is given in Schwarzschild coordinates and does not cover the full spacetime, but only one of the four patches at a time. In order to set up the functional,
it is therefore more convenient to work in Eddington--Finkelstein coordinates which cover two adjacent regions in the same chart, e.g., regions I and II. In these coordinates, the metric
reads
\begin{equation}
    ds^2 = \frac{1}{z^2}\left(-\rho e^{-\chi}dv^2-2e^{-\chi/2}dv dz + \sum_{i=1}^{d-1}dx_i^2\right)
    \quad \text{with} \quad
    v = t-\int^z_0 dz' \frac{e^{\chi/2}}{\rho},
\end{equation}
and the boundary points that the half surface is connecting are given by
\begin{equation}
    \left(v_\text{I} = \frac{\tau}{2}, \, z_\text{I} = 0 \right) \quad \text{and} \quad \left(v_\text{II} = - \int_{0}^{z_\text{max}} dz' \frac{e^{\chi/2}}{\rho}, \, z_\text{II} = z_\text{max}\right),
\end{equation}
with $z_\text{max}$ fixed by the smoothness condition at the gluing point, as explained above.
\subsection{Case 1: Embedding-independent Functionals}
In the case where $\zeta$ does not depend on the embedding, we can write $\zeta(X^\mu, g_{\mu\nu}) = \zeta(g_{\mu\nu}) = a(z)$ since the metric components depend only on $z$. Pulling back the metric onto
the surface, the functional takes the explicit form
\begin{equation}
    F_a[\Sigma] = 2 z_\text{h}^{d-1} A_\text{h}\int d\lambda z^{-d}\sqrt{- \rho e^{-\chi}\dot{v}^2 - 2e^{-\chi/2}\dot{v}\dot{z}}a(z),
\end{equation}
where $A_\text{h}$ is the (infinite) horizon area obtained by integrating over the planar directions defined in Equation~\eqref{eq:bekenstein_hawking_entropy}. In the following, we take
the parameter $\lambda$ to satisfy
\begin{equation}
    \sqrt{-\rho e^{-\chi}\dot{v}^2-2e^{-\chi/2}\dot{v}\dot{z}} = z^{-d+2}e^{-\chi/2}a(z),
    \label{eq:parameterization_scalar}
\end{equation}
which is unique up to a sign. The sign is chosen by imposing $\dot{z} < 0$, i.e., the half-surface starts in region II and moves towards the boundary in region I.
The equation of motion describing the extremal surface is then given by
\begin{equation}
    \frac{\dot{z}^2}{z^4} = -\mathcal{U}(\pi_v, z) = \pi_v^2 - U(z)
    \quad \text{with} \quad
    U(z) = -\rho(z)e^{-\chi(z)}a^2(z)z^{-2d},
    \label{eq:eom_embedding}
\end{equation}
where $\pi_v = (-\dot{z} - \rho e^{-\chi/2} \dot{v})/z^2$ is the conjugate momentum associated to the coordinate $v$, which is conserved along the extremal surface since $v$ is cyclic. 
The smoothness condition at the gluing point translates into $\dot{z}|_{z_\text{max}} = 0$, which is equivalent to $\mathcal{U}(\pi_v, z_\text{max}) = 0$. 

As mentioned above, the observable $\mathcal{O}_a$ depends on the sum of the boundary times $\tau$. Using standard techniques from variational calculus, it is possible to show that 
the growth rate of the observable with respect to $\tau$ is given by
\begin{equation}
    \frac{d\mathcal{O}_a}{d\tau} = z_\text{h}^{d-1} A_\text{h} \pi_v.
    \label{eq:complexity_growth_rate}
\end{equation}
Had we chosen an orientation for $\lambda$ such that the half surface starts in region I and moves towards the gluing point in region II, i.e., $\dot{z} > 0$, we would get a minus 
sign in front of the expression on the right-hand side.

Although $\pi_v$ is conserved along the extremal surface, it still depends on $\tau$, since hypersurfaces anchored at different boundary times carry different momenta. 
The relationship between $\pi_v$ and $\tau$ is obtained from
\begin{equation}
    \tau = 2 v_\text{I} = 2\left(\int_{v_\text{II}}^{v_\text{I}} dv+v_\text{II}\right)
    = -2\int_0^{z_\text{max}} dz \frac{\rho \dot{v}+e^{\chi/2}\dot{z}}{\rho \dot{z}} 
    = -2\pi_v \int_0^{z_\text{max}} dz \frac{e^{\chi/2}}{\rho \sqrt{-\mathcal{U}}},
    \label{eq:tau_momentum_relation}
\end{equation}
where we used the definition of $\pi_v$ and the negative branch of Equation~\eqref{eq:eom_embedding}. To study the late-time growth of $\mathcal{O}_a$, we
focus on extremal surfaces with large $\tau$, denoted $\Sigma_\infty$ in Figure~\ref{fig:extremal_surface}. The corresponding momentum $\pi_\infty$ can then be obtained from the condition
that the integral above diverges positively.

As shown in Appendix~\ref{sec:a_boundary_contributions}, the contributions from the $z=0$ boundary to the $\tau$-integral are finite for all examples that we consider.
Since the divergence at the horizon vanishes in the PV prescription, the divergence must occur at the upper bound. Smoothness ensures that the integrand diverges at $z_\text{max}$, but 
for generic values of $z_\text{max}$ and $\pi_v$, this divergence is integrable.
A non-integrable divergence arises only if $\mathcal{U}(\pi_v, z)$ develops a local maximum at $z_\text{max} = z_\infty$. Hence, the late- and early-time parameters $\pi_{\pm\infty}$ and $z_{\pm\infty}$
satisfy
\begin{equation}
    \mathcal{U}(\pi_{\pm\infty}, z_{\pm\infty}) = 0,
    \quad
    \mathcal{U}'(\pi_{\pm\infty}, z_{\pm\infty}) = 0,
    \quad \text{and} \quad
    \mathcal{U}''(\pi_{\pm\infty}, z_{\pm\infty}) < 0.
    \label{eq:critical_conditions}
\end{equation}
In the present case, the dependence of $\mathcal{U}$ on $\pi_v$ is particularly
simple, so that these conditions reduce to the requirement that $U(z)$ has a maximum behind the horizon, with $\pi_{\pm\infty} = \pm\sqrt{U(z_{\pm\infty})}$ and $z_{\infty}=z_{-\infty}$. Thus, an extremal surface at late times
only exists if such a maximum exists. When it does, the observable grows linearly with $\tau$ at late times as expected for complexity observables, with growth rate determined by $\pi_\infty$.

As an example of a function $\zeta$ that does not depend on the embedding but only on geometric invariants in the ambient space, we consider the case
\begin{equation}
    \begin{aligned}
        a(z) &= 1+\frac{\lambda_{C^2}}{d(d-1)^2(d-2)}C_{\mu\nu\rho\sigma}C^{\mu\nu\rho\sigma} \\
        &=1+\frac{\lambda_{C^2}}{4d^2(d-1)^2}z^4(\rho\chi'^2-2\rho\chi''-3\chi'\rho'+2\rho'')^2,
        \label{eq:a_weyl}
    \end{aligned}
\end{equation}
where $C_{\mu\nu\rho\sigma}$ is the Weyl tensor of the background spacetime and $\lambda_{C^2}$ is a coupling constant. For the hairless AdS black hole,
this expression reduces to $a(z) = 1 + \lambda_{C^2} u^{2d}$, which implies that the effective potential of Equation~\eqref{eq:eom_embedding} takes the form
\begin{equation}
    U(z) = \left(u^{-d}-u^{-2d}\right)\left(1+\lambda_{C^2}u^{2d}\right)^2.
    \label{eq:potential_weyl}
\end{equation}
\begin{figure}[!htbp]
    \centering
    \includegraphics[width=0.95\textwidth]{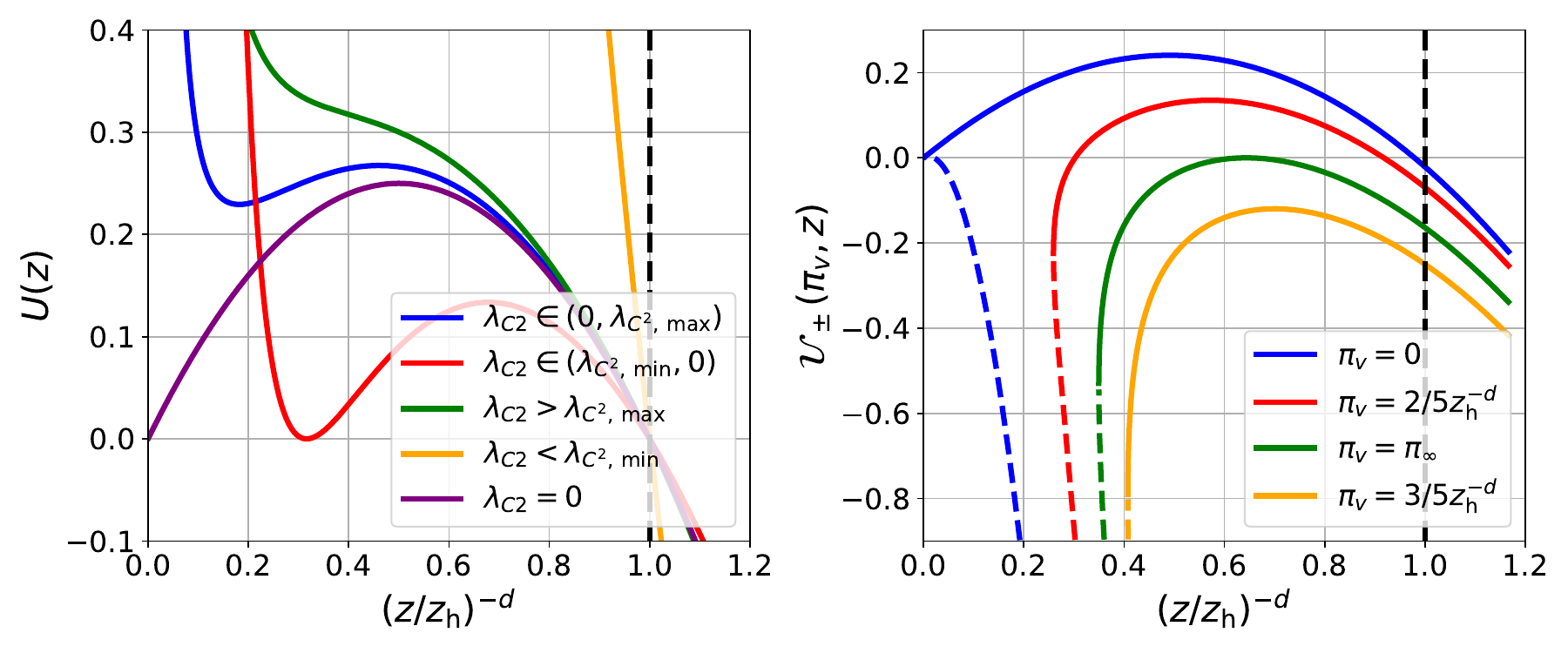}
    \caption{Left: Potential of the $C^2$-functional from Equation~\eqref{eq:potential_weyl} for various values of $\lambda_{C^2}$. A local maximum behind the horizon appears only when 
    $\lambda_{C^2}$ lies within the critical range. Right: Effective potentials $\mathcal{U}_{\pm}(\pi_v, z)$ corresponding to the $K$-functional of 
    Equation~\eqref{eq:effective_potential_extrinsic}, shown for $\lambda_K = 1/10$ and several choices of $\pi_v$. Solid curves denote the negative branch, while dashed curves indicate 
    the positive branch. Among these, only the negative branch evaluated at $\pi_\infty$ satisfies the late-time conditions. The left panel applies to any dimension $d$, whereas the 
    right panel is specific to $d = 3$. The vertical line marks the location of the horizon.}
    \label{fig:potential}
\end{figure}
A plot of the potential for different values of $\lambda_{C^2}$ is shown in the left panel of Figure~\ref{fig:potential}. As indicated in the figure, $U(z)$ only develops a 
local maximum behind the horizon if the parameter $\lambda_{C^2}$ obeys
\begin{equation}
    -1 = \lambda_{C^2, \, \text{min}} < \lambda_{C^2} < \lambda_{C^2, \, \text{max}} = \frac{47-13\sqrt{13}}{8} \approx 0.01598.
    \label{eq:lambda_bounds}
\end{equation}
An analysis of the behavior outside this critical range is provided in Appendix~\ref{sec:b_late_time}.
For all $\lambda_{C^2}$ within this range, the location of the maximum is given by
\begin{equation}
    \begin{aligned}
    u_\infty^{-d} =& \frac{1}{6}\Biggl(1+\frac{1+12\lambda_{C^2}}{\left(1-144\lambda_{C^2}+6\sqrt{3}\sqrt{-\lambda_{C^2}(4\lambda_{C^2}(4\lambda_{C^2}-47)+3)}\right)^{1/3}} \\
    &+\left(1-144\lambda_{C^2}+6\sqrt{3}\sqrt{-\lambda_{C^2}(4\lambda_{C^2}(4\lambda_{C^2}-47)+3)}\right)^{1/3}\Biggr)
    \end{aligned}
    \label{eq:zinfty_weyl}
\end{equation}
with $u_\infty = z_\infty/z_\text{h}$. Due to the finite range of $\lambda_{C^2}$, 
$z_\infty$ cannot become arbitrarily big, but takes its maximal value at the upper bound of $\lambda_{C^2}$.
The conserved momentum at late times associated
to the observable's growth rate can be computed by taking the square root of $U(z_\infty)$, leading to
\begin{equation}
    \pi_\infty = \frac{4\left(1-u_\infty^{-d}\right)^{3/2} u_\infty^{-d/2}}{3-2u_\infty^{-d}}z_\text{h}^{-d}.
    \label{eq:pi_infty_weyl}
\end{equation}
Just as $z_\infty$, the value of $\pi_\infty$ is bounded and becomes maximal at the upper bound of $\lambda_{C^2}$.

\subsection{Case 2: Embedding-dependent Functionals}
The case where $\zeta$ depends on the embedding is more complicated and cannot be treated as generally as the previous case. We will therefore focus on the specific example 
\begin{equation}
    \zeta_K(X^\mu, g_{\mu\nu}) = 1+\lambda_K K,
    \label{eq:zeta_extrinsic}
\end{equation}
where $K$ is the mean curvature of the surface (given by the trace of the extrinsic curvature $K_{\mu\nu}$) and $\lambda_K$ is again a coupling constant. By definition, $K$ depends on the
second derivatives of the embedding functions, so that the functional now contains second derivatives with respect to the parameter $\lambda$. However, it can be shown that the 
higher-derivative terms are total derivatives that do not contribute to the equations of motion, which remain second order. In principle we would also need to add boundary terms to have a well-posed variational problem, but they would not contribute to the equations of motion either, so we do not need to worry about them
for the following discussion. The explicit form of the extrinsic curvature of the surfaces of interest in the spacetime with metric given by Equation~\eqref{eq:ansatz} reads
\begin{equation}
    \begin{aligned}
        K =& \frac{e^{-3\chi/2}}{2\left(-\rho e^{-\chi}\dot{v}^2-2e^{-\chi/2}\dot{z}\dot{v}\right)^{3/2}} \Bigl(z\rho \rho'\dot{v}^3 - \rho^2\dot{v}^3\left(2d + z\chi'\right)
        - e^{\chi} \dot{z}^2 \dot{v}\left(4d + z\chi'\right) \\
        & + 2e^{\chi} z\dot{v} \ddot{z} + e^{\chi/2}\dot{z}\Bigl(-3\rho\dot{v}^2\left(2d + z\chi'\right) 
        + z\Bigl(3\rho'\dot{v}^2 - 2e^{\chi/2} \ddot{v}\Bigr)\Bigr)\Bigr).
    \end{aligned}
    \label{eq:mean_curvature}
\end{equation}
Inserting this back into the functional, it follows that the conserved momentum associated to the cyclic variable $v$ is given by
\begin{equation}
    \pi_v = -\frac{1}{z^2}\left(\dot{z}+\rho e^{-\chi/2}\dot{v}\right)
    -\lambda_K\frac{e^{-\chi/2}}{z^d}\left(\frac{\rho'z}{2}-\rho\left(d+\frac{z\chi'}{2}\right)\right)
    -(d-1)\lambda_K e^{\chi/2}z^{d-4} \dot{z}^2,
\end{equation}
where the parameterization of Equation~\eqref{eq:parameterization_scalar} has again been used with $a(z)=1$. Equation~\eqref{eq:complexity_growth_rate} holds also in the case where 
$\zeta$ depends on the embedding, so $\pi_v$ still determines the growth rate of the observable with respect to $\tau$. With the explicit expression for $\pi_v$ at hand, it is possible to 
derive the equation of motion for the $z$-coordinate, resulting in
\begin{equation}
    \begin{aligned}
    \frac{\dot{z}^2}{z^4} =& \frac{e^{-\chi}z^{-2d}}{2 (d-1)^2 \lambda_K^2} \Biggl(1 - 2(d-1) \pi_v \lambda_K e^{\chi/2}z^d + 2(d-1)\lambda_K^2 \left(d\rho-\frac{z}{2}(\rho'-\rho\chi') \right) \\
    & \pm  \sqrt{
         1 - 4(d-1) \pi_v  \lambda_K e^{\chi/2}z^d
        +4(d-1)\lambda_K^2\left(\rho-\frac{z}{2}(\rho'-\rho\chi')\right)
    }\Biggr) \\
    =& -\mathcal{U}_{\pm}(\pi_v, z).
    \label{eq:effective_potential_extrinsic}
\end{aligned}
\end{equation}
The minus branch reduces to the pure volume case for $\lambda_K\to 0$, while the plus branch diverges in this limit. A plot of the effective potential for both branches is shown in the right
panel of Figure~\ref{fig:potential}. For the present case, the relationship between $\tau$ and $\pi_v$ becomes
\begin{equation}
    \begin{aligned}
        \tau =& -2 \int_0^{z_\text{max}} dz
        \frac{e^{\chi/2}}{\rho \sqrt{-\mathcal{U}_{\pm}}} 
        \Biggl(
            \pi_v+\lambda_K \Biggl(
            \frac{e^{-\chi/2}}{z^d} \Bigl(\frac{\rho'z}{2}-\rho \Bigl(d + \frac{z\chi'}{2} \Bigr)\Bigr) \\
        &-(d-1) e^{\chi/2} z^d \mathcal{U}_{\pm}
            \Biggr)
        \Biggr).
    \end{aligned}
    \label{eq:tau_momentum_relation_extrinsic}
\end{equation}
As shown in Appendix~\ref{sec:a_boundary_contributions}, the integrand remains finite as $z\to 0$ for both branches of the potential if the background spacetime asymptotes to AdS.
Therefore, in order to find extremal surfaces at late times, the conditions in Equation~\eqref{eq:critical_conditions} need to be solved again.
For the hairless AdS black hole, the effective potential reads
\begin{equation}
    \begin{aligned}
        \mathcal{U}_{\pm}(\pi_v, z) =& -\frac{1}{2(d-1)^2\lambda_K^2}\Biggl(u^{-2d}-2\lambda_K(d-1)\pi_v u^{-d}+
        \lambda_K^2(d-1)u^{-2d}\bigl(2d \\
        &-d u^d\bigr)
        \pm u^{-2d}\sqrt{1-4\lambda_K\left(d-1\right)\pi_v u^d+2\lambda_K^2(d-1)
        \left(2+(d-2)u^d\right)}\Biggr).
        \label{eq:potential_K}
    \end{aligned}
\end{equation}
As explained in Appendix~\ref{sec:c_time_evolution_k}, the late- and early time conditions can only be satisfied for the negative branch, with the critical values given by
\begin{equation}
    u_{\pm\infty}^{-d} = \frac{1}{2}\left(1\pm\frac{\lambda_K d}{\sqrt{1+\lambda_K^2 d^2}}\right)
    \quad \text{with} \quad
    \pi_{\pm\infty} = \pm\frac{1}{2}\sqrt{1+\lambda_K^2 d^2}z_\text{h}^{-d}.
    \label{eq:pi_infty_extrinsic}
\end{equation}
In contrast to the $C^2$-observable discussed previously, there is no restriction on the coupling constant $\lambda_K$ for the existence of extremal surfaces at late times.
\section{Singularity Probes}
\label{sec:probes}

In the previous section, we introduced two different CAny observables and analyzed their late-time behavior in the background of a hairless AdS black hole. We now want to go
further and study how these observables behave in the presence of scalar hair, first for the Chamblin--Reall solution and then for the massive scalar solution. 
For both scenarios, we will work with $d=3$.

\subsection{Proper Time to Approach the Singularity}
The closest approach of the extremal surfaces to the singularity is given by the turning point $z_\text{max}$. However, since this quantity is coordinate dependent, it does
not provide a robust measure of proximity. A more physically meaningful alternative is to consider the proper time experienced by an observer travelling along a geodesic from $z_\text{max}$
to the singularity at $z \to \infty$. The proper time for a not necessarily geodesic timelike curve is given by
\begin{equation}\label{tau_proper}
    \tau_\text{proper} = \int d\lambda \frac{1}{z}\sqrt{\rho e^{-\chi} \dot{t}^2 - \frac{1}{\rho} \dot{z}^2},
\end{equation}
where we chose Schwarzschild coordinates again since we are only considering region II of the Penrose diagram in Figure~\ref{fig:extremal_surface}. Note that the sign in
the square root has changed compared to the previous section since we are now dealing with timelike curves. As before, $\lambda$ is an arbitrary parameter along the trajectory,
defined to point towards the future, i.e., the singularity, and the dot denotes a derivative with respect to it. We fix $\lambda$ to be the proper time itself, such that we have
\begin{equation}
    \frac{1}{z}\sqrt{\rho e^{-\chi} \dot{t}^2 - \frac{1}{\rho} \dot{z}^2} = 1.
\end{equation}
The coordinate $t$ is cyclic again, leading to a conserved momentum $E_t = 1/z^2 \rho e^{-\chi} \dot{t}$. By algebraically solving the proper time constraint for $\dot{z}$, we find
the radial equation of motion
\begin{equation}
    \dot{z}^2 = E_t^2 z^4 e^\chi-\rho z^2.
\end{equation}
Plugging the results back in Eq.~\eqref{tau_proper}, we find that for a geodesic with conserved momentum $E_t$, the proper time it takes to fall from $z_\text{max}$
behind the horizon into the singularity is given by
\begin{equation}
    \tau_\text{max} = \int_{z_\text{max}}^{\infty} \frac{dz}{\sqrt{E_t^2 z^4 e^\chi-\rho z^2}}.
\end{equation}
To obtain a unique value, we fix $E_t = 0$, which corresponds to an observer initially
at rest at the horizon, and then falling into the black hole. This observer will eventually cross the extremal surface at $z_\text{max}$, and the proper time then measures the
time it takes for this observer to continue from $z_\text{max}$ to the singularity. One can verify that this is also the largest possible time it takes to fall from $z_\text{max}$
to the singularity. For the hairless AdS black hole, the proper time can be computed analytically, yielding
\begin{equation}
    \tau_\text{max} = \frac{2}{d}\arctan\left(\frac{1}{\sqrt{z_\text{max}^d-1}}\right).
\end{equation}
In all other cases, the integral has to be solved numerically. The proper time from the turning point of the extremal late-time surface to the singularity will be denoted by
$\tau_\infty$.

\subsection{Chamblin--Reall Background}
We start by analyzing the CAny observables in the Chamblin--Reall background. Previously, we have seen that in order to modify the values of the Kasner exponents, we 
need to change the value of the parameter $\alpha$. As discussed, the value of $\phi_\text{h}$ needs to be chosen as in Equation~\eqref{eq:phi_h} for the asymptotic form of the metric
to remain fixed. Since both the $C^2$-observable and the $K$-observable depend on a parameter $\lambda_{C^2}$ or $\lambda_K$, we will scan
the two-dimensional parameter space spanned by $\alpha$ and $\lambda_{C^2}$ or $\lambda_K$ for both observables to determine for which values of the parameters extremal late-time surfaces exist, how
close these surfaces get to the singularity, and what the late-time growth rate of the observables is. In the previous section, this has been done analytically for the case $\alpha=0$.
We will numerically extend these results to $\alpha > 0$ here.

\subsubsection{\boldmath $C^2$-Observable}
The $C^2$-observable is characterized by the function $a(z)$ defined in Equation~\eqref{eq:a_weyl}. Plugging the Chamblin--Reall solution given in Equations~\eqref{eq:cr_solutions} 
into the expression for $a(z)$, we obtain
\begin{equation}
    a(z) = 1+\lambda_{C^2} e^{2\alpha \phi_h} u^{2d+\eta}.
    \label{eq:a_cr}
\end{equation}
Substituting this into the form of the potential displayed in Equation~\eqref{eq:eom_embedding} and eliminating $\phi_\text{h}$ with Equation~\eqref{eq:phi_h}, one obtains
\begin{equation}
    U(z) = -z^{-2d}\left(1+\lambda_{C^2}\left(1-\frac{\eta}{2d}\right)^2 u^{2d+\eta}\right)^2\left(1-u^{d\left(1-\eta/(2d)\right)}\right).
\end{equation}
In order to have extremal surfaces that reach late times, the potential needs to have a local maximum behind the horizon. We have seen that for $\alpha=0$, this is the case only if
$\lambda$ lies within the bounds given in Equation~\eqref{eq:lambda_bounds}, with the location of the maximum $z_\infty$ and the corresponding conserved momentum $\pi_\infty$ given 
in Equations~\eqref{eq:zinfty_weyl} and \eqref{eq:pi_infty_weyl}, respectively.

\begin{figure}[!htbp]
    \centering
    \includegraphics[width=0.95\textwidth]{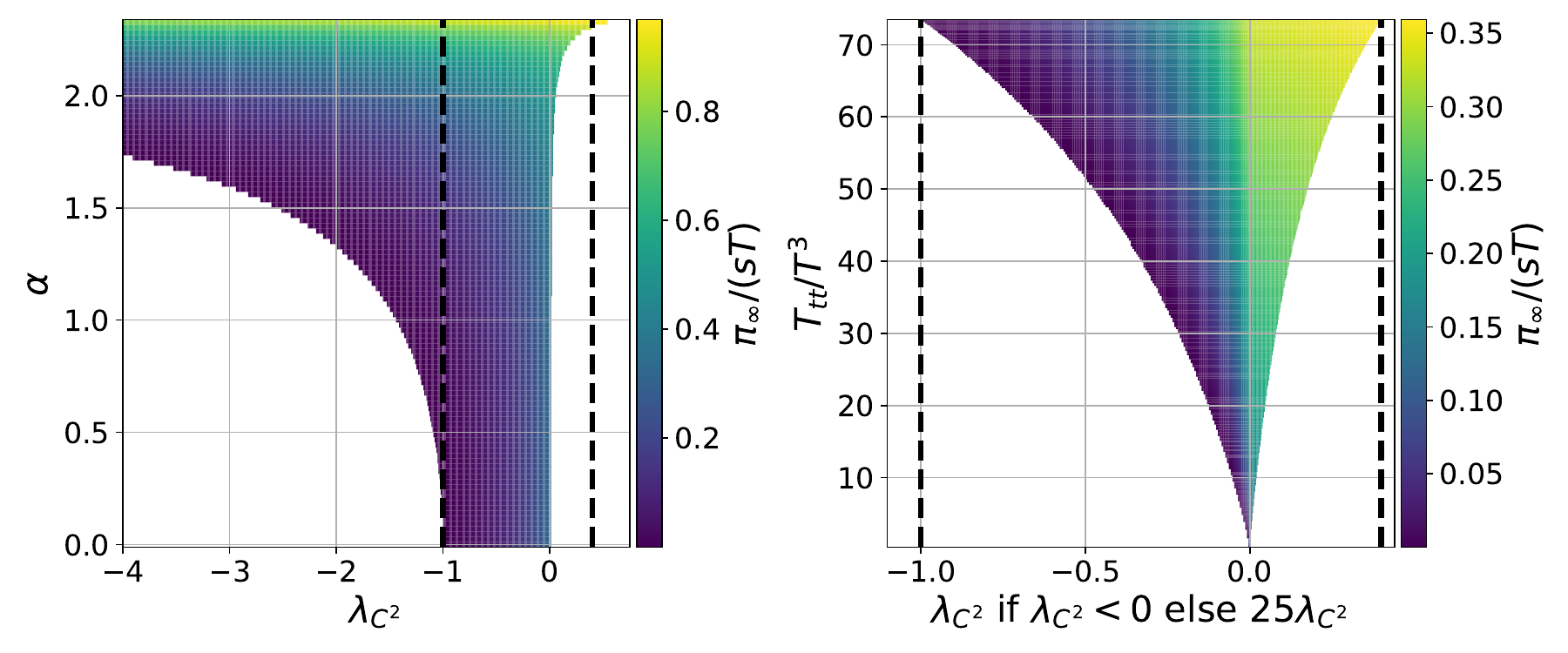}
    \caption{Parameter scan for the $C^2$-observable in the Chamblin--Reall background (left) and the massive scalar background (right). The shaded region indicates the parameter values for which the potential
    has a local maximum behind the horizon, implying the existence of extremal surfaces anchored at late times. The corresponding value of $\pi_\infty/(sT)$ is shown by the color.
    The dashed lines mark the analytic bounds on $\lambda_{C^2}$ obtained for the hairless black hole.}
    \label{fig:c2_scan}
\end{figure}
For $\alpha>0$, we numerically check for which parameter combinations of $\alpha$ and $\lambda_{C^2}$ a local maximum behind the horizon exists. The result is shown in the left panel of 
Figure~\ref{fig:c2_scan}. 
The shaded area indicates the region where a local maximum exists behind the horizon, 
with the color representing the value of $\pi_\infty/(sT)$. The normalization by the product of the entropy density $s$ and the temperature $T$ has been chosen to make the growth rate
dimensionless and not depend on the location of the horizon $z_\text{h}$. The numerical results reproduce the analytic bounds on $\lambda_{C^2}$ at $\alpha=0$ represented by the vertical 
dashed lines. As $\alpha$ increases, the allowed interval for $\lambda_{C^2}$ expands in both directions. 

\begin{figure}[!htbp]
    \centering
    \includegraphics[width=0.95\textwidth]{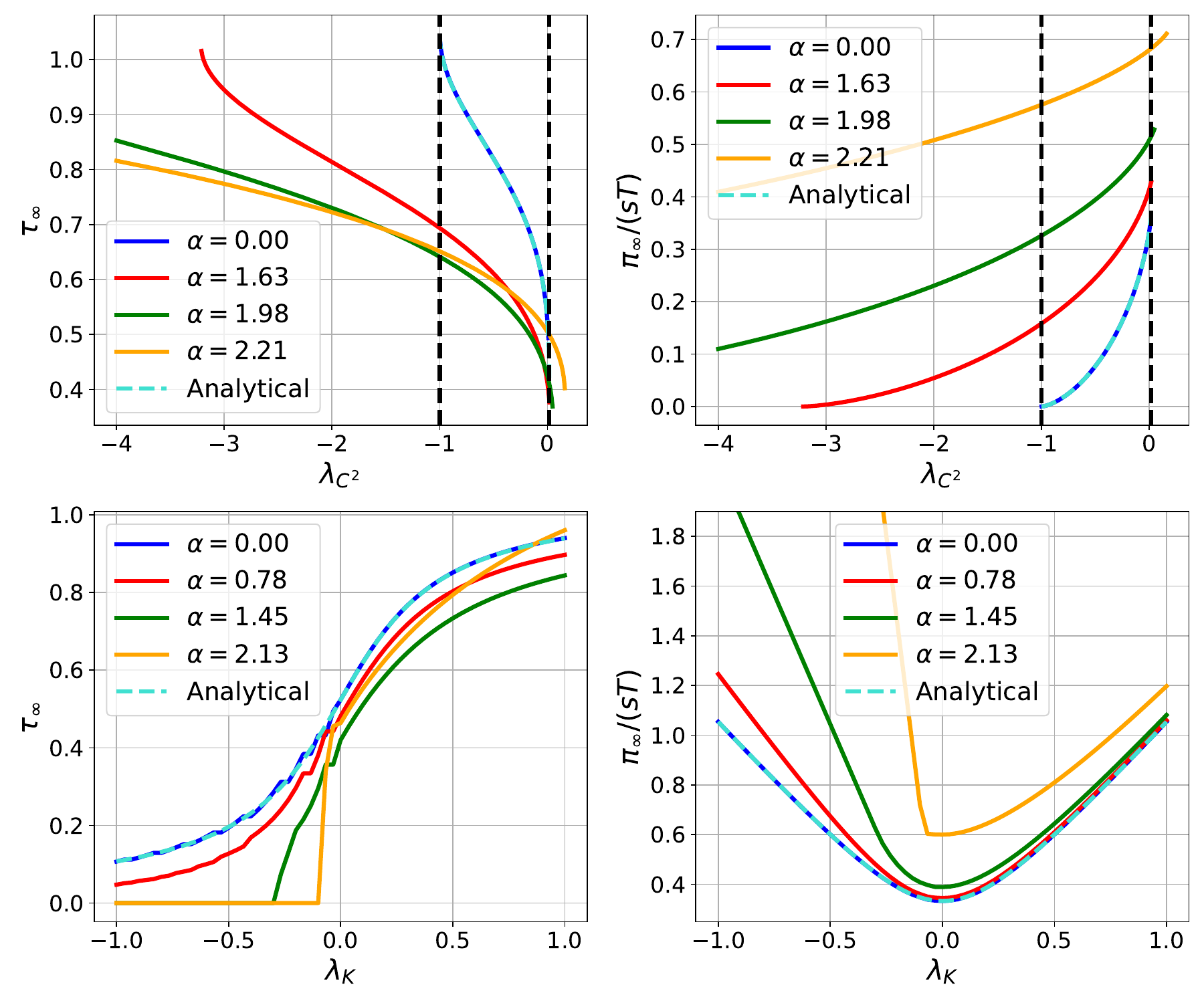}
    \caption{Proper time $\tau_\infty$ from $z_\infty$ to the singularity (left column) and late-time growth rate 
    $\pi_\infty/(sT)$ (right column) for the $C^2$-observable (top row) and the $K$-observable (bottom row) in the 
    Chamblin--Reall background. All quantities are shown as functions of the coupling $\lambda_{C^2}$ or $\lambda_K$ for 
    several values of $\alpha$. For the $C^2$-observable, the allowed range of $\lambda_{C^2}$ is indicated by vertical 
    dashed lines. Turquoise curves denote analytic results at $\alpha=0$.}
    \label{fig:cr_slices}
\end{figure}

The upper row of Figure~\ref{fig:cr_slices} shows the proper time $\tau_\infty$ (left) and the late-time growth 
rate $\pi_\infty/(sT)$ (right) as functions of $\lambda_{C^2}$ for several values of $\alpha$. Because the upper bound
on $\lambda_{C^2}$ remains finite for all $\alpha$, $\tau_\infty$ cannot become arbitrarily small. Consequently, extremal $C^2$-surfaces never reach the singularity. For all values of $\alpha$, $\tau_\infty$ 
decreases monotonically with $\lambda_{C^2}$. The dependence on $\alpha$ is monotonic as well for fixed $\lambda_{C^2}$, 
but the direction of the effect depends on the value of $\lambda_{C^2}$.
The right plot shows that $\pi_\infty/(sT)$ grows monotonically with $\lambda_{C^2}$. This means that, within a fixed 
background, late-time surfaces that extend closer to the singularity also exhibit a larger growth rate. Moreover, 
increasing $\alpha$ raises $\pi_\infty/(sT)$ for all values of $\lambda_{C^2}$.

\subsubsection{\boldmath $K$-Observable}
We now turn to the $K$-observable. Plugging the Chamblin--Reall solution into the expression for its associated
effective potential given in Equation~\eqref{eq:effective_potential_extrinsic} yields
\begin{equation}
    \begin{aligned}
        \mathcal U_{\pm}(\pi_v,z)
        =&
        -\frac{
        z^{-2d}u^{-\eta}
        }{
        8(d-1)^{2}\lambda_K^{2}
        }
        \Biggl(4
        +(d-1)\lambda_K u^{\eta/2}
        \Bigl(
        8d\lambda_K u^{\eta/2}
        - \lambda_K(4d
        +(d-1)\alpha^2)u^{d} \\
        &- 8\pi_v z^{d}
        \Bigr)
        \pm 2\Bigl(4+2(d-1)\lambda_K u^{\eta/2}
        \Big(8\lambda_K u^{\eta/2}-\lambda_K
        (8-4d \\
        &+(d-1)\alpha^2)u^{d}
        -8\pi_v z^{d}
        \Big)\Bigr)^{1/2}
        \Biggr).
    \end{aligned}
\end{equation}
As explained in Appendix~\ref{sec:a_boundary_contributions}, the integrand of the $\tau$-integral determining the anchoring time remains finite at the conformal boundary for all values of 
$\alpha$ for the negative branch. For the positive branch, however, finiteness requires
\begin{equation}
    \frac{\alpha^2}{4}(d-1) < 1.
\end{equation}
If this condition is violated, the integral generically diverges to $-\infty$ for $\lambda_K > 0$ and to $+\infty$ for $\lambda_K < 0$. When the
condition is satisfied, the positive branch is unable to produce solutions for late or early times: we found that for $\alpha=0$, it fails to meet the late-time conditions in 
Equation~\eqref{eq:critical_conditions}, and numerical tests for $\alpha>0$ suggest that this remains true even for hairy solutions. Thus, the positive branch either admits extremal
surfaces only at finite $\tau$ if the above condition holds, or only at infinite $\tau$ if it does not. We are primarily interested in the late-time
limit. If the above condition is violated and $\lambda_K < 0$, extremal surfaces obtained with the positive branch might
give a larger value of the observable than those obtained with the negative branch. If this is the case, the positive branch
becomes physically relevant and there is a discontinuous jump in complexity as the anchoring time goes to infinity.
In the following, we will still focus on the negative branch, but should keep this caveat in mind. 

For $\alpha=0$, the negative branch develops a local maximum behind the horizon for any value of $\lambda_K$, with the corresponding 
turning point $z_\infty$ and conserved momentum $\pi_\infty$ given in Equation~\eqref{eq:pi_infty_extrinsic}. A numerical scan confirms 
that this persists for all tested values of $\alpha$. The resulting behavior is shown in the lower row of 
Figure~\ref{fig:cr_slices}. The left panel illustrates that for sufficiently negative $\lambda_K$, the turning point can be pushed 
arbitrarily close to the singularity, allowing the $K$-observable to probe it. Increasing $\alpha$ moves the surfaces even deeper into 
the interior in this range of $\lambda_K$. 
The right panel shows that for $\alpha=0$, $\pi_\infty/(sT)$ is symmetric in $\lambda_K$. As $\alpha$ increases, this 
symmetry is progressively broken: negative $\lambda_K$ for which the surface is more penetrating yield larger growth rates. In contrast to the $C^2$-observable, surfaces
with smaller $\tau_\infty$ at a fixed background geometry do not generally give larger values of $\pi_\infty/(sT)$.

\subsection{Massive Scalar Background}
For the massive scalar background, the Kasner exponents can be modified by changing the value of the scalar field at the horizon $\phi_\text{h}$, or equivalently, the value
of $\langle T_{tt}\rangle/T^3$, as shown in Figure~\ref{fig:Kasner_Exponents}. In addition to $d=3$, we fix the mass of the scalar field to $m^2=-2$ for the numerical analysis.

\subsubsection{\boldmath $C^2$-Observable}
The parameter scan over $\langle T_{tt}\rangle/T^3$ and $\lambda_{C^2}$ for extremal late-time $C^2$-surfaces is shown in the right panel of
Figure~\ref{fig:c2_scan}. Unlike in the Chamblin--Reall background, where the allowed interval of $\lambda_{C^2}$ widens as $\alpha$ increases, here the opposite occurs: the admissible
range of $\lambda_{C^2}$ shrinks as the normalized energy density decreases. For sufficiently small $\langle T_{tt}\rangle/T^3$, the only choice that admits extremal late-time
surfaces is the volume functional, $\lambda_{C^2}=0$.

\begin{figure}[!htbp]
    \centering
    \includegraphics[width=0.95\textwidth]{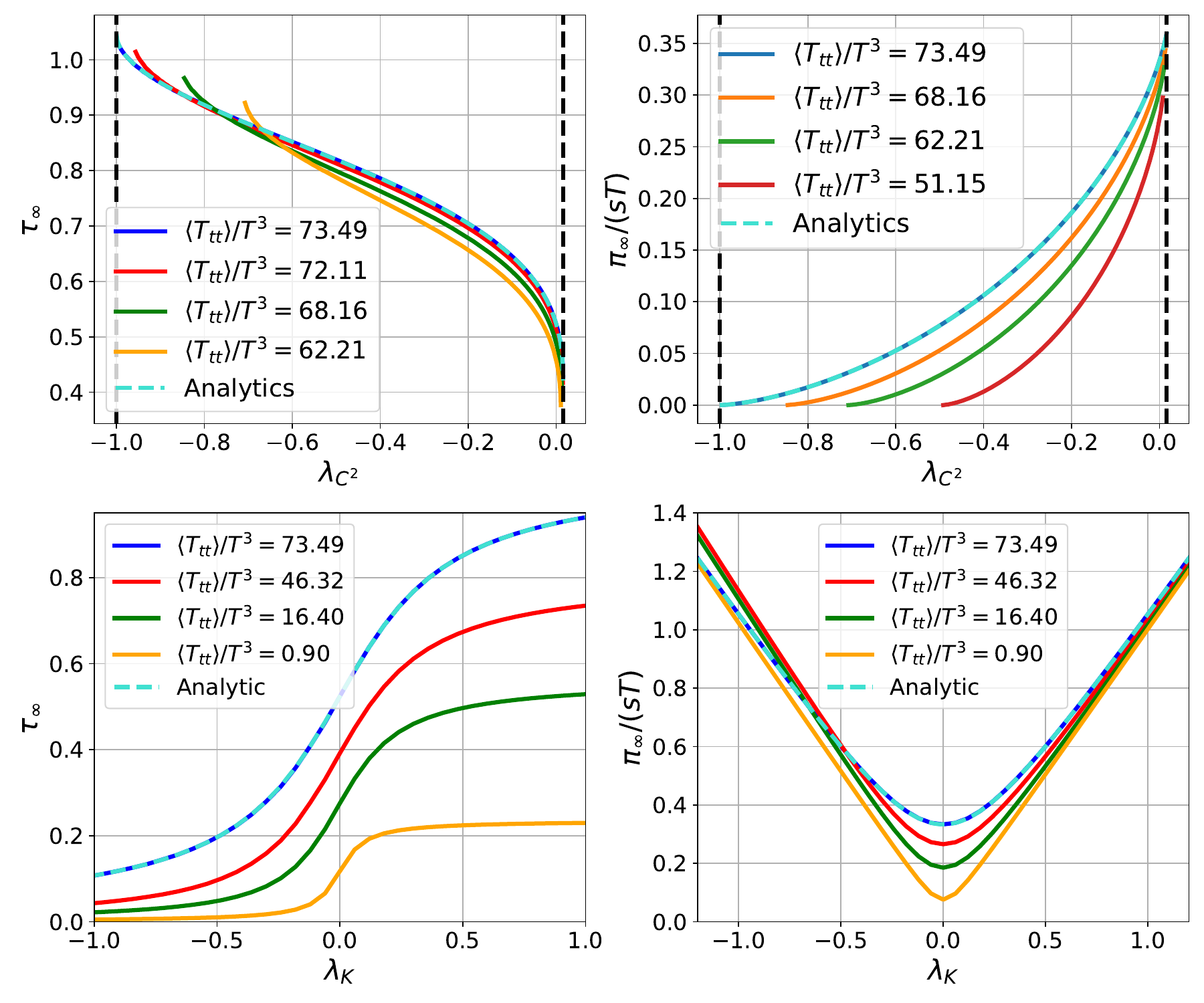}
    \caption{Proper time $\tau_\infty$ from $z_\infty$ to the singularity (left column) and late-time growth rate
    $\pi_\infty/(sT)$ (right column) for the $C^2$-observable (top row) and the $K$-observable (bottom row) in the 
    massive scalar background. All quantities are shown as functions of the coupling $\lambda_{C^2}$ or $\lambda_K$ for 
    several values of $\langle T_{tt}\rangle/T^3$. For the $C^2$-observable, the allowed range of $\lambda_{C^2}$ is indicated by vertical 
    dashed lines. Turquoise curves denote analytic results at $\langle T_{tt}\rangle/T^3=64\pi^3/27$.}
    \label{fig:massive_slices}
\end{figure}

The corresponding behavior of $\tau_\infty$ and $\pi_\infty/(sT)$ is displayed in the upper row of Figure~\ref{fig:massive_slices}. As in the Chamblin--Reall case, 
$\tau_\infty$ decreases monotonically with $\lambda_{C^2}$ but remains bounded below due to the finite upper limit on $\lambda_{C^2}$. 
Its dependence on $\langle T_{tt}\rangle/T^3$ is again monotonic at fixed $\lambda_{C^2}$, with the direction of the change determined by the value of $\lambda_{C^2}$. 
For a fixed background, surfaces that probe more deeply into the interior continue to exhibit larger growth rates. 
However, in contrast to the dependence on $\alpha$ in the Chamblin--Reall geometry, we find that $\pi_\infty/(sT)$ now decreases monotonically with 
$\langle T_{tt}\rangle/T^3$ for all values of $\lambda_{C^2}$.

\subsubsection{\boldmath $K$-Observable}
As in the Chamblin--Reall background, the numerics indicate that the negative branch of the potential admits
extremal late-time surfaces for any value of $\lambda_K$ throughout the full range of
$\langle T_{tt} \rangle / T^3$. The resulting behavior is shown in the lower row of
Figure~\ref{fig:massive_slices}. The left panel illustrates that decreasing
$\langle T_{tt}\rangle/T^3$ pushes the turning point $z_\infty$ closer to the singularity for
all $\lambda_K$.
The right panel shows the corresponding late-time growth rate. The symmetry in $\lambda_K$ for the hairless
black hole is again broken, but the asymmetry is considerably weaker here. Moreover, the dependence of $\pi_\infty/(sT)$ on
$\langle T_{tt} \rangle / T^3$ becomes negligible at large $|\lambda_K|$.
Because the massive scalar background approaches AdS at the boundary, the anchoring time of the
positive branch remains finite for all parameter values. Thus, the discontinuous late-time transition that can potentially
arise in the Chamblin--Reall background does not occur here.

\subsection{Probing the Kasner Exponents}
We have found that the $C^2$-observable is unable to probe the singularity in either background. In contrast, the $K$-observable is well suited
for accessing the singularity, and its sensitivity improves when $\alpha$ is 
increased in the Chamblin--Reall background or when $\langle T_{tt} \rangle / T^3$ is decreased in the massive scalar background. Since the geometry near the singularity is 
characterized by the Kasner exponents, it is natural to investigate how the $K$-observable responds to variations in these parameters. 

\begin{figure}[!htbp]
    \centering
    \includegraphics[width=0.95\textwidth]{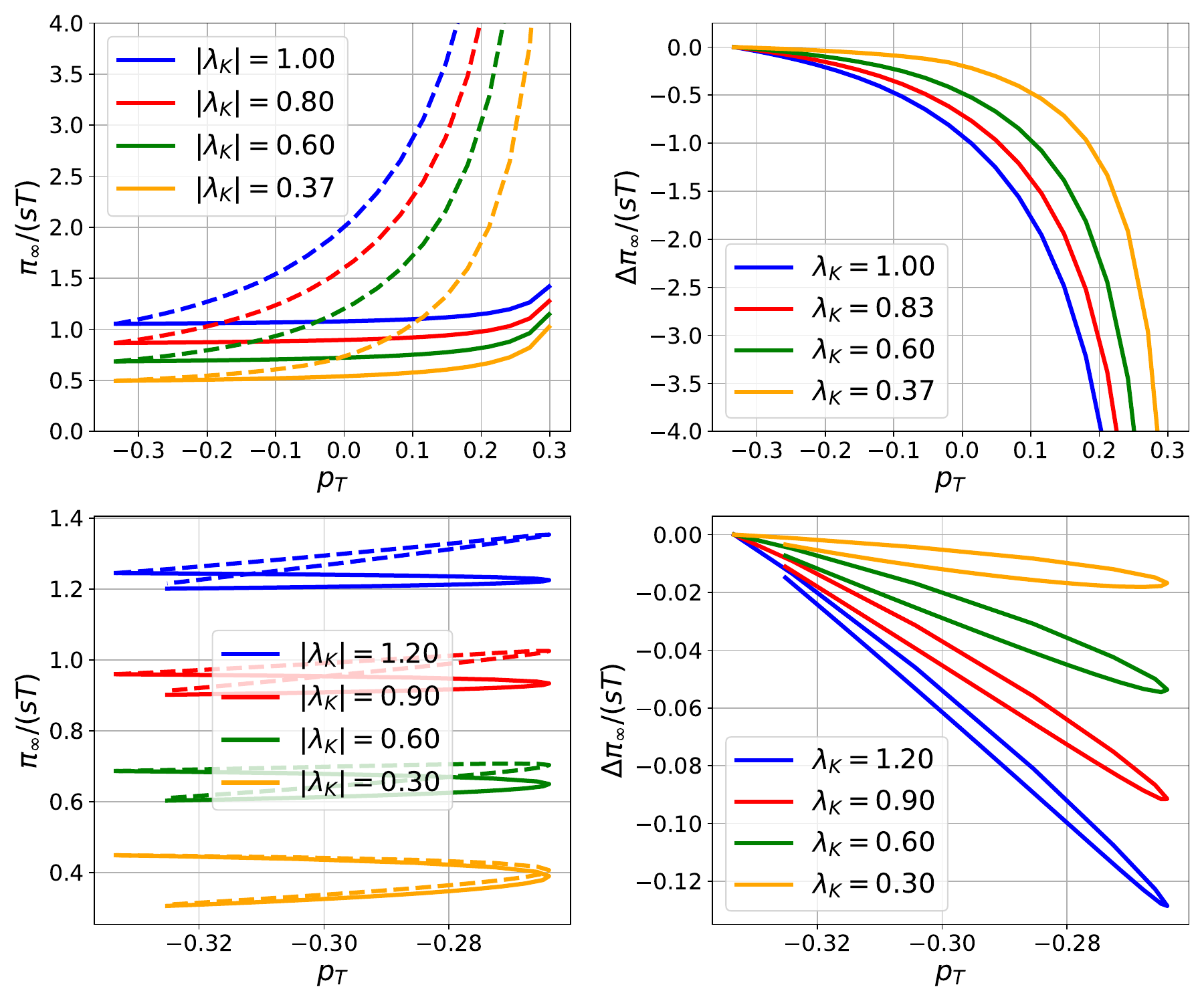}
    \caption{Investigating the asymmetry of $\pi_\infty/(sT)$ with respect to the sign of $\lambda_K$ observed in Figures~\ref{fig:cr_slices} and \ref{fig:massive_slices}. The left column
    shows $\pi_\infty/(sT)$ as a function of the Kasner exponent $p_T$ for several values of $|\lambda_K|$, with solid lines corresponding to positive $\lambda_K$ and dashed lines to negative
    $\lambda_K$. The right column displays the asymmetry $\Delta \pi_\infty/(sT)$ defined in Equation~\eqref{eq:asymmetry} again as a function of $p_T$ for several values of $\lambda_K$. 
    The top row corresponds to the Chamblin--Reall background and the bottom row to the massive scalar background.}
    \label{fig:kasner_probes}
\end{figure}

A central observation is that $\pi_\infty/(sT)$ is symmetric in $\lambda_K \to -\lambda_K$ for the hairless black hole,
while this symmetry is broken when scalar hair is introduced. Figure~\ref{fig:kasner_probes} examines this effect in detail. The left column shows $\pi_\infty/(sT)$ as a function of the
Kasner exponent $p_T$ for different values of $|\lambda_K|$, with solid curves representing positive $\lambda_K$ and dashed curves negative $\lambda_K$. The right column 
illustrates the corresponding asymmetry $\Delta \pi_\infty/(sT)$ defined as
\begin{equation}
    \Delta \pi_\infty = \pi_\infty(\lambda_K) - \pi_\infty(-\lambda_K),
    \label{eq:asymmetry}
\end{equation}
again plotted against $p_T$. For the Chamblin--Reall background (top row), the negative $\lambda_K$ curves grow more rapidly with $p_T$ than the positive $\lambda_K$ curves. As expected, 
the symmetry is restored at the vacuum value $p_T = -1/3$,  where the two curves coincide and the asymmetry vanishes. Away from this point, the asymmetry grows, 
with negative $\lambda_K$ surfaces, which probe deeper into the Kasner regime, yielding larger values of $\pi_\infty/(sT)$. 
Because $p_T$ increases monotonically with $\alpha$ (see Figure~\ref{fig:kasner_cr}),
each value of $p_T$ corresponds to a unique $\pi_\infty/(sT)$.

The massive scalar background (bottom row) exhibits a similar qualitative trend, but with an important difference: the map between $p_T$ and $\langle T_{tt}\rangle/T^3$ is
non-monotonic (see Figure~\ref{fig:Kasner_Exponents}), so there are always two values of $\pi_\infty/(sT)$ for a given $p_T$. Even so, the negative $\lambda_K$ curves consistently 
grow more rapidly with $p_T$ than the positive $\lambda_K$ curves, again reflecting that surfaces with negative $\lambda_K$ are more penetrating.

In the Chamblin--Reall background, the asymmetry is roughly an order of magnitude larger than in the massive scalar background, as can be seen by comparing the vertical scales in the corresponding plots. This enhancement can be traced to the behavior of the Kasner exponents: in the Chamblin--Reall geometry they vary significantly more across the parameter range than in the masive scalar case. This larger dynamical range, visible on the right-hand sides of Figures~\ref{fig:kasner_cr} and \ref{fig:Kasner_Exponents}, naturally amplifies the difference between positive and negative $\lambda_K$, and hence leads to a more pronounced asymmetry in $\pi_\infty/(sT)$.


\section{Conclusion}
\label{sec:conclusion}

In this work, we investigated the behavior of two distinct CAny observables in the backgrounds of two different scalar-haired black holes.
The first geometry features a scalar field governed by an exponential potential and admits the analytic Chamblin–Reall solution \cite{Chamblin:1999ya}, while the second involves 
a purely massive scalar field and requires numerical integration. In both settings, the presence of scalar hair alters the near-singularity geometry,
allowing the associated Kasner exponents to deviate continuously from their vacuum values.

The two codimension-one CAny observables we studied include one depending on the Weyl tensor of the ambient spacetime ($C^2$-observable) and one depending
on the extrinsic curvature of an embedded surface ($K$-observable). Both observables are parameterized by a coupling constant that smoothly interpolates between the given observable and
the volume functional.
We found that the $C^2$-observable admits extremal late-time surfaces only within a finite interval of the coupling constant. This interval grows when the scalar hair in the 
Chamblin--Reall background is made more influential, but shrinks in the massive-scalar background. Because of the boundedness of the interval, the corresponding extremal late-time 
surfaces cannot be pushed arbitrarily close to the singularity, indicating that the $C^2$-functional is unable to probe the deepest interior regions of either black hole. In addition, the fact that the region of parameters admitting linear growth shrinks to zero for low-temperature massive-scalar black holes suggests that this functional may not be a good complexity measure in general.

In contrast, the $K$-observable admits extremal late-time surfaces for all values of its coupling constant in both backgrounds. This freedom allows the extremal surfaces to approach 
the singularity arbitrarily closely. Moreover, in the regime where the late-time surfaces probe the near-singularity region, strengthening the impact of the scalar hair drives the 
surfaces even deeper into the interior. This behavior shows that the $K$-functional provides a sensitive and tunable probe of the near-singularity geometry in scalar-haired black holes.

Finally, while the late-time growth rate of the $K$-observable is symmetric under sign flips of its coupling constant in hairless black holes, this symmetry is broken once scalar hair is
introduced. We observed this asymmetry in both the Chamblin--Reall and the massive scalar backgrounds, with negative coupling enhancing the penetration depth of the surfaces and thereby 
increasing the growth rate. The effect becomes most pronounced when the Kasner exponent $p_T$ approaches its maximal value.

As a next step, the numerical analysis in the massive-scalar background could be extended by introducing a gauge field, following \cite{Auzzi:2022bfd,Yang:2019gce}.
Without the scalar, this setup reduces to the Reissner–Nordström black hole, whose inner Cauchy horizon prevents CAny observables from probing arbitrarily deep into the interior \cite{Jorstad:2023kmq}.
Once the scalar is switched on, however, the Cauchy horizon becomes unstable and is replaced by a spacelike singularity \cite{Hartnoll:2020rwq}, potentially reopening access to the deep interior and offering a new setting in which to test the observables studied here.

A complementary direction is to explore different complexity functionals. One possibility is to use two distinct functionals: one to determine the extremal surface and another to evaluate the complexity.
This idea was implemented in \cite{Jorstad:2023kmq}, where several functionals were evaluated on constant mean curvature (CMC) slices obtained as extremal surfaces of a codimension-zero functional.
Since the resulting observables exhibited markedly different behavior near the singularity, this approach may help to isolate features more directly correlated with the Kasner exponents.
Another extension that might establish a sharper relationship between the observables and the near-singularity geometry is to consider subleading corrections to the late-time growth rate. Indeed it may be interesting to analyze the full time dependence of the complexity and not only the late-time growth rate.

Finally, our analysis of the two observables in the Chamblin–Reall background revealed that their late-time growth rates increase as the parameter $\alpha$ approaches its upper bound.
This behavior contrasts with that of hydrodynamic quantities, which instead slow down in this regime \cite{Gursoy:2015nza,Betzios:2017dol,Betzios:2018kwn}.
Clarifying the origin of this difference could provide an interesting new perspective on the interplay of the dynamics of complexity and other observables.

\begin{acknowledgments}
We thank R. Auzzi, S. Bolognesi, S. Chapman, R.C. Myers and E. Rabinovici for useful comments, and S.A. Hartnoll for sharing his code, that we used in the initial stages of the project. 
\end{acknowledgments}

\appendix
\section{Boundary Contributions in Chamblin--Reall Geometries}
\label{sec:a_boundary_contributions}

In this appendix, we analyze the boundary contributions to the $\tau$-integral relating the anchoring time and the conserved momentum of extremal surfaces, as well as to the values of 
the CAny observables in the Chamblin--Reall background.

\subsection{Anchoring Time}
For the $C^2$-functional where $\zeta$ does not depend on the embedding, the $\tau$-integral takes the form
\begin{equation}
    \tau = \int_0^{z_\text{max}} dz f_{C^2}(z),
\end{equation}
where the explicit expression for the integrand is given by the right hand side of Equation~\eqref{eq:tau_momentum_relation}.
For values of $\alpha$ within the allowed range, blackening function of the Chamblin--Reall solution asymptotes to
\begin{equation}
        \rho(z) \to \beta u^{\eta}
        \quad \text{with} \quad
        \beta = \frac{e^{\alpha\phi_h}}{1-\eta/(2d)}
    \label{eq:asymptotics_cr}
\end{equation}
near the boundary. Substituting this together with the form of $\chi$ given in Equation~\eqref{eq:cr_solutions} into 
$f_{C^2}(z)$, we obtain the asymptotic behavior
\begin{equation}
    f_{C^2}(z)
    \to
    -2\pi_v \beta^{-3/2} z_\text{h}^{\eta/2} z^{d-\eta/2},
\end{equation}
where Equation~\eqref{eq:a_cr} has been used to take $a(z)\to 1$. Because the exponent is positive for all admissible $\alpha$, the integrand vanishes as $z\to 0$. Thus, the
$\tau$-integral for the $C^2$-functional receives a finite contribution from the lower bound.
For the $K$-functional defined in Equation~\eqref{eq:zeta_extrinsic}, the $\tau$-integral reads
\begin{equation}
    \tau = \int_0^{z_\text{max}} dz f_{K,\pm}(z),
\end{equation}
with the integrand given in Equation~\eqref{eq:tau_momentum_relation_extrinsic} and $\mathcal{U}_\pm$ defined in 
Equation~\eqref{eq:effective_potential_extrinsic}. For $z \to 0$, the two branches of the effective potential scale as
\begin{equation}
    \mathcal{U}_\pm \to 
    \begin{cases}
        &-\frac{z_\text{h}^\eta}{(d-1)^2\lambda_K^2} z^{-2d-\eta}, \\
        &-\beta z^{-2d},
    \end{cases}
\end{equation}
where the upper line corresponds to the positive branch and the lower line to the negative branch. This implies that $f_{K,\pm}(z)$ behaves asymptotically as
\begin{equation}
    f_{K, \pm}(z) \to
    \begin{cases}   
        &-\frac{2\sigma(\lambda_K)}{\beta} u^{-\eta/2}, \\
        &\frac{2\lambda_K}{\sqrt{\beta}},
    \end{cases}
\end{equation}
where $\sigma(\lambda_K)$ denotes the sign of $\lambda_K$.
The result shows that the negative branch always yields a finite contribution at the lower limit, whereas the positive branch does so only when $\alpha=0$.
After performing the integration, the positive branch contributes a finite value at the lower bound only if
\begin{equation}
    \eta/2 < 1
\end{equation}
is satisfied.
If this bound is violated, which is possible within the allowed range of $\alpha$, the integral diverges to $-\infty$ for $\lambda_K > 0$ and to
$+\infty$ for $\lambda_K < 0$. Consequently, in this regime, the positive branch admits extremal
surfaces to the $K$-functional only in the limit of infinite anchoring time. In contrast, the negative branch is free from
this restriction and supports extremal surfaces for all values of $\tau$.

\subsection{Observables}
In this section, the boundary contribution of the Chamblin--Reall solution to the holographic complexity observables considered in the main 
text will be analyzed. These contributions are obtained by evaluating the corresponding functionals on extremal surfaces. We start with the $C^2$-observable, which takes the form
\begin{equation}
    \mathcal{O}_a = \int_0^{z_\text{max}} dz \underbrace{2 z_\text{h}^{d-1} A_h\frac{z^{-2d}e^{-\chi/2}}{\sqrt{\pi_v^2 - U(z)}}a(z)^2}_{=g_{C^2}(z)}.
\end{equation}
It is easy to show that asymptotically, the integrand behaves as
\begin{equation}
    g_{C^2}(z) \to \frac{2 z_\text{h}^{d+\eta/2-1} A_h}{\sqrt{\beta}} z^{-d-\eta/2},
\end{equation}
which implies that the boundary contribution depends on $\alpha$ and deviates from the AdS case with $\alpha=0$. The value of the $K$-observable reads
\begin{equation}
    \mathcal{O}_K = \int_0^{z_\text{max}} dz \underbrace{2 z_\text{h}^{d-1} A_h\frac{z^{-2d}e^{-\chi/2}}{\sqrt{-\mathcal{U}_\pm}}(1+\lambda_K K)}_{=g_{K,\pm}(z)},
\end{equation}
where we again made use of the gauge fixing condition given in Equation~\eqref{eq:parameterization_scalar}, using $a(z)=1$. The expression for the mean curvature $K$ is given by
\begin{equation}
    \begin{aligned}
        K = \frac{z^{3d-6}}{2} \Bigl(& z\rho \rho'\dot{v}^3 - \rho^2\dot{v}^3\left(2d + z\chi'\right)
        - e^{\chi} \dot{z}^2 \dot{v}\left(4d + z\chi'\right) + 2e^{\chi} z\dot{v} \ddot{z} \\
        & + e^{\chi/2}\dot{z}\left(-3\rho\dot{v}^2\left(2d + z\chi'\right) 
        + z\left(3\rho'\dot{v}^2 - 2e^{\chi/2} \ddot{v}\right) \right)
        \Bigr).
    \end{aligned}
\end{equation}
To determine its scaling behavior near the boundary, we first need the asymptotics of $\dot{v}$ and $\dot{z}$. Using the gauge fixing condition, we solve for $\dot{v}$, yielding
\begin{equation}
    \dot{v} = -\frac{e^{\chi/2}}{\rho}\dot{z} \pm
    \sqrt{\frac{e^\chi}{\rho^2}\dot{z}^2 - \frac{1}{\rho}z^{-2d+4}}.
\end{equation}
The expression contains a sign ambiguity in front of the square root. In our convention, the half surface starts at the center in region II and moves toward the boundary in region I, 
implying that $\dot{z} < 0$ always. This direction also forces that $v$ increases monotonically, i.e. $\dot{v} > 0$. In the expression above, the first term
may become positive or negative, depending on the sign of $\rho$ which is positive in region I and negative in region II. Therefore, the only possible way to ensure that $\dot{v} > 0$ is to pick the positive sign for the square root.
To verify that picking the positive sign indeed guarantees $\dot{v} > 0$, we need to show that
\begin{equation}
     \sqrt{\frac{e^\chi}{\rho^2}\dot{z}^2 - \frac{1}{\rho}z^{-2d+4}} > \frac{e^{\chi/2}}{\rho}\dot{z}.
\end{equation}
The left-hand side is always positive. Since the right-hand side is negative in region I due to $\rho > 0$ and $\dot{z} < 0$,
the relation is always satisfied in region I. In region II, the right-hand side is positive as well as $\rho < 0$. Squaring on both sides, we end up 
with the relation
\begin{equation}
    -\frac{1}{\rho} z^{-2d+4} > 0,
\end{equation}
which is also always true in region II. Therefore, picking the positive sign for the square root indeed ensures that $\dot{v} > 0$ along the entire surface.
Using Equation~\ref{eq:effective_potential_extrinsic}, it follows that 
the asymptotic form of $\dot{z}$ reads
\begin{equation}
    \dot{z} = -z^2\sqrt{-\mathcal{U}_\pm}
    \to
    \begin{cases}
         -\frac{z_\text{h}^{\eta/2}}{(d-1)|\lambda_K|}z^{2-d-\eta/2}, \\
        -\sqrt{\beta} z^{2-d}.
    \end{cases}
\end{equation}
Plugging this result back into the expression for $\dot{v}$, it follows that to leading order, $\dot{v}$ scales as
\begin{equation}
    \dot{v} \to
    \begin{cases}
        \frac{2 z_\text{h}^\eta}{(d-1)|\lambda_K|\beta} z^{2-d-\eta}, \\
        \frac{z_\text{h}^{\eta/2}}{\sqrt{\beta}} z^{2-d-\eta/2}
    \end{cases}
\end{equation}
close to the boundary. With the asymptotic forms of $\dot{z}$ and $\dot{v}$ at hand, one can find that the mean curvature
behaves as
\begin{equation}
    K \to
    \begin{cases}
        \frac{4d}{(d-1)^3|\lambda_K|^3\beta}u^{-\eta}, \\
        -\frac{\eta\sqrt{\beta}}{2}u^{\eta/2}.
    \end{cases}
\end{equation}
We can plug this result back into the integrand of the functional to find its asymptotic behavior. After some algebra, we find
\begin{equation}
    g_{K, \pm}(z) \to
    \begin{cases}
        \frac{8d z_\text{h}^{d+\eta-1} A_\text{h}}{(d-1)^2\lambda_K\beta} z^{-d-\eta}, \\
        \frac{2z_\text{h}^{d+\eta/2-1} A_\text{h}}{\sqrt{\beta}} z^{-d-\eta/2}.
    \end{cases}
\end{equation}
In both cases, the degree of divergence of the functional at the boundary depends on $\alpha$, so just as for the $C^2$-functional,
the Chamblin--Reall solution gives different boundary contributions than AdS to the $K$-functional.

\section{\boldmath Time Evolution of the $C^2$-Observable}
\label{sec:b_late_time}

The CAny observables discussed in Section~\ref{sec:complexity} where $\zeta$ does not depend on the embedding admit extremal surfaces at late times if the associated potential given in 
Equation~\eqref{eq:eom_embedding} has a local maximum behind the horizon. As an example, we specialized in the $C^2$-observable, whose associated potential is defined in 
Equation~\eqref{eq:potential_weyl}. We found that the potential only develops a sub-horizon maximum if the coupling constant $\lambda_{C^2}$ lies within the range given in 
Equation~\eqref{eq:lambda_bounds}, as shown in Figure~\ref{fig:potential}.

Solving Equation~\eqref{eq:eom_embedding} can be viewed as determining the trajectory of a particle with energy $\pi_v^2$ moving in the potential $U(z)$. The turning point of the surface 
coincides with the classical turning point of the particle, where it momentarily comes to rest before rolling back down the potential. The trajectory corresponding to the late-time 
surface is the critical one in which the particle climbs the potential and comes to rest exactly at its maximum. The associated energy is given by $\pi_\infty^2$.

\begin{figure}[!htbp]
    \centering
    \includegraphics[width=0.95\textwidth]{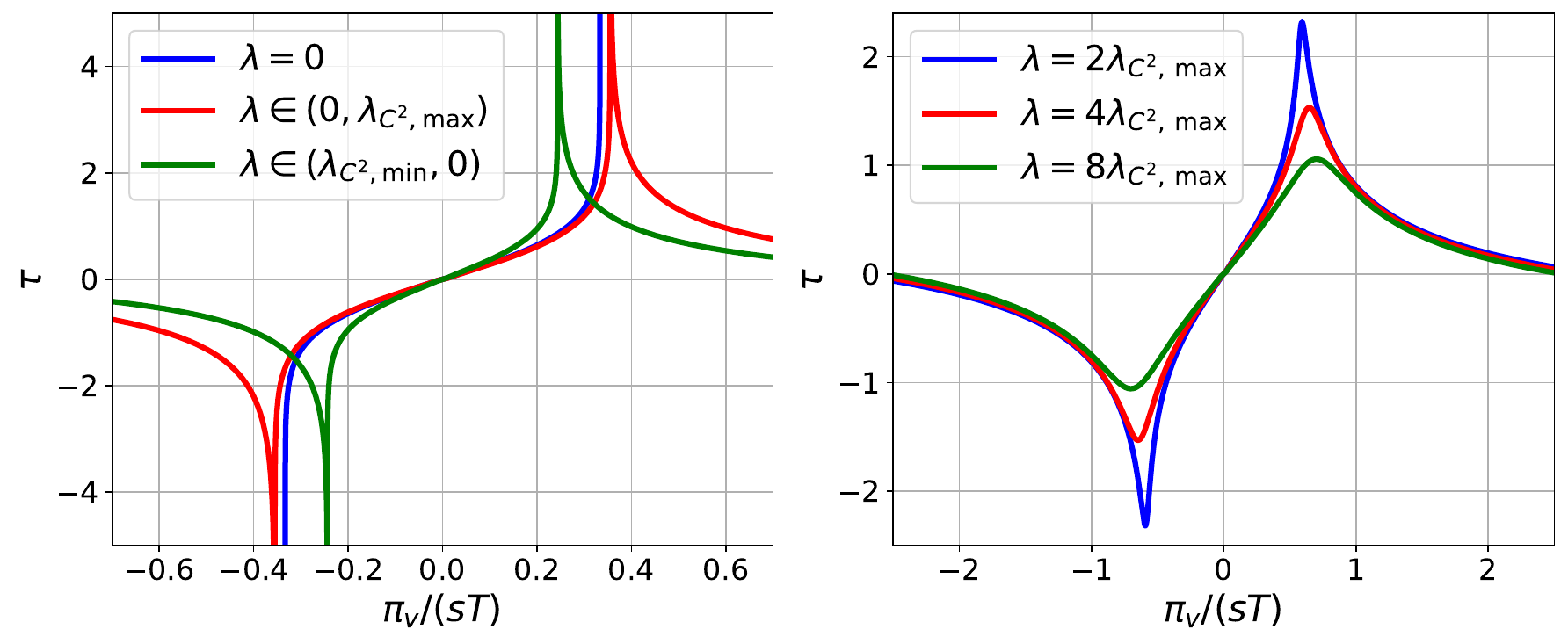}
    \caption{Anchoring time $\tau$ as a function of the conserved momentum $\pi_v$. Left: For values of $\lambda_{C^2}$ within the critical range, the effective potential has a local maximum behind the horizon. This implies that $\tau$ can become infinite for $\pi_\infty$. Right: If $\lambda_{C^2}$ lies outside the critical range, the potential has no such maximum, and $\tau$ remains finite for all $\pi_v$. In this regime, no extremal surfaces exist at late times.}
    \label{fig:tau_pi_relation}
\end{figure}

Figure~\ref{fig:potential} shows that in the volume case $\lambda_{C^2}=0$, the potential is bounded from above and vanishes at the singularity. This excludes solutions with 
$\pi_v^2 > \pi_\infty^2$. For $\lambda_{C^2}\neq 0$, however, the potential diverges near the singularity, implying that solutions exist for all energies. Consequently, $\pi_v$ can 
approach $\pi_\infty$ either from below or from above. This behavior is shown in the left plot of Figure~\ref{fig:tau_pi_relation}. The authors of \cite{Belin:2022xmt} have shown that 
in the background of hairless AdS black holes, the \enquote{dipping branch} in which the particle shoots over the maximum and then rolls back, always yields a smaller value of $F_{C^2}$ 
than the branch approaching $\pi_\infty$ from below. Therefore, only the latter is relevant for determining the complexity.

When $\lambda_{C^2}$ does not lie within the critical range, the effective potential has no local maximum behind the horizon. In this case, it is natural to ask how $\tau$ behaves as 
$\pi_v$ is increases and the particle probes deeper into the black hole interior. This dependence is shown in the right panel of Figure~\ref{fig:tau_pi_relation}. The curves reveal 
that $\tau$ reaches a maximum for a finite value of $\pi_v$. Therefore, multiple values of $\pi_v$ correspond to the same anchoring time. The close parallel to the behavior in the 
critical range suggests that, among the possible branches, the one with the smaller momentum again yields the larger contribution to the holographic complexity.

\section{\boldmath Time Evolution of the $K$-Observable}
\label{sec:c_time_evolution_k}

\begin{figure}[!htbp]
    \centering
    \includegraphics[width=0.95\textwidth]{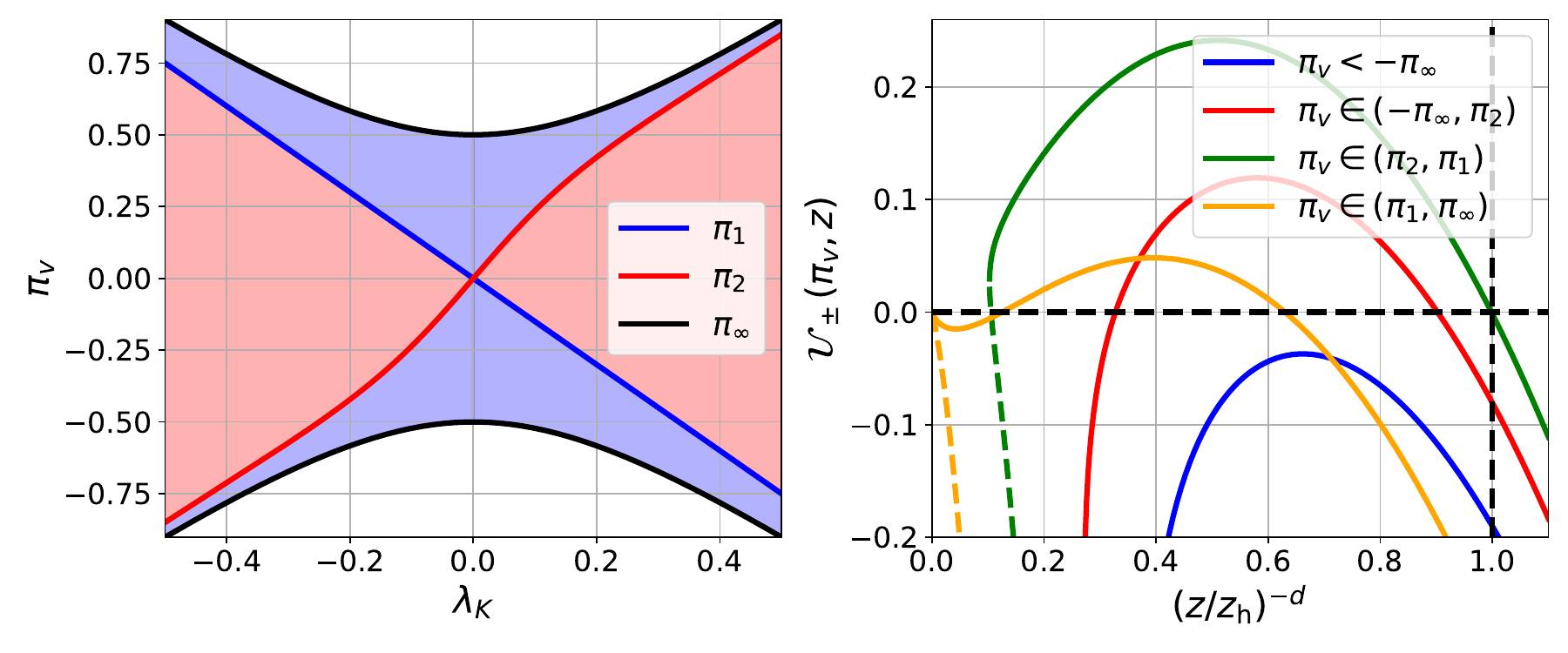}
    \caption{Left: Depending on the value of $\pi_v$, the $u_-$ root defined in Equation~\eqref{eq:roots} belongs to the negative (blue region) or positive branch (red region) of the 
    potential. If $\pi_v$ lies outside of the critical range, the root disappears (white region). Right: Effective potential for $d=3$ and $\lambda_K = -1/10$ at different values of 
    $\pi_v$, illustrating the different cases. Only for the orange curve, Equation~\eqref{eq:pi_crit} is satisfied, such that the root at the singularity appears.    
    The full lines represent the negative branch, while the dashed lines correspond to the positive branch.}
    \label{fig:root_belonging}
\end{figure}

The effective potential associated with the $K$-observable in the background of an AdS black hole is given in Equation~\eqref{eq:potential_K}. We have seen that two branches for the 
potential exist, both of which give rise to extremal surfaces. The two branches can potentially have roots at the singularity $u^{-d}=0$ and at
\begin{equation}
    u_\pm^{-d}
    =\frac{1}{2(1+\lambda_K^2 d^2)}
    \left(1+\lambda_K^2 d^2+2\lambda_K d\pi_v r_\text{h}^{-d}
    \pm\sqrt{1+\lambda_K^2 d^2-4\pi_v^2 r_\text{h}^{-2d}}\right).
    \label{eq:roots}
\end{equation}
The root at the singularity belongs to both branches, but only appears if the argument of the square root appearing in the potential remains positive up to this point, which requires 
\begin{equation}
    \pi_v
        \begin{cases}
        \le \pi_{\text{crit}} & \text{if $\lambda_K > 0$}, \\
        \ge \pi_{\text{crit}} & \text{if $\lambda_K < 0$},
        \end{cases}
    \quad \text{with} \quad
    \pi_{\text{crit}} = \frac{d-2}{2}\,\lambda_K r_h^d.
    \label{eq:pi_crit}
\end{equation}
The $u_\pm$ roots merge for $\pi_v=\pm\pi_\infty$ with $\pi_\infty$ defined in Equation~\eqref{eq:pi_infty_extrinsic}. If $\pi_v$ does not satisfy the constraint
\begin{equation}
    -\pi_\infty \leq \pi_v \leq \pi_\infty,
    \label{eq:pi_constraint}
\end{equation}
the two roots become complex and no longer lie on the real axis. It can be shown that the $u_+$ root always belongs to the negative branch, while the $u_-$ root can potentially belong
to either branch. The critical values of $\pi_v$ separating the two cases are given by
\begin{equation}
    \pi_1 = -d\lambda_K \frac{r_\text{h}^d}{2}
    \quad \text{and} \quad
    \pi_2 =
    \sigma(\lambda_K)
    \sqrt{
    1 + d^2 \lambda_K^2 - 
    \left( \frac{4 d \lambda_K^2 - 4 \lambda_K^2 + 1}{4 d^2 \lambda_K^2 - 8 d \lambda_K^2 + 4 \lambda_K^2 + 1} \right)^2
    }\frac{r_h^d}{2},
\end{equation}
where $\sigma(\lambda_K)$ denotes the sign of $\lambda_K$. The $u_-$ root belongs to the positive branch if
\begin{equation}
    \min(\pi_1, \pi_2) \leq \pi_v \leq \max(\pi_1, \pi_2),
\end{equation}
and to the negative branch in all other cases, as long as Equation~\eqref{eq:pi_constraint} is satisfied. This is visualized in the left panel of Figure~\ref{fig:root_belonging}, with the 
region shaded in blue/red representing the regime where the $u_-$ root belongs to the negative/positive branch, respectively. The right panel of Figure~\ref{fig:root_belonging} shows 
the effective potentials for the different cases.

To find extremal surfaces at late or early times, the conditions in Equation~\eqref{eq:critical_conditions} need to be satisfied, which can only happen if two roots merge. Since the 
$u_+$ root always belongs to the negative branch if it is real, only the negative branch can have late-time extremal surfaces that do not crash into the singularity. 
For the positive branch, Equation~\eqref{eq:critical_conditions} can only be satisfied if $u_-$ merges with the root at the singularity, which happens for $\pi_v=\pi_1$.
However, this still fails to produce extremal surfaces at late or early times since it can be shown that this particular choice of $\pi_v$ kills the divergent mode of the integrand in
Equation~\eqref{eq:tau_momentum_relation_extrinsic}.
Therefore, we conclude that the positive branch does not produce additional extremal surfaces at late times. At finite times, several extremal surfaces can exist due to the 
double branching, and to tell which one is associated to complexity, one has to figure out for which surface the $K$-functional becomes larger. Since in the main text we are only 
investigating the late-time limit, we will not dive deeper into this analysis here.

\begin{figure}[htbp!]
    \centering
    \includegraphics[width=0.95\textwidth]{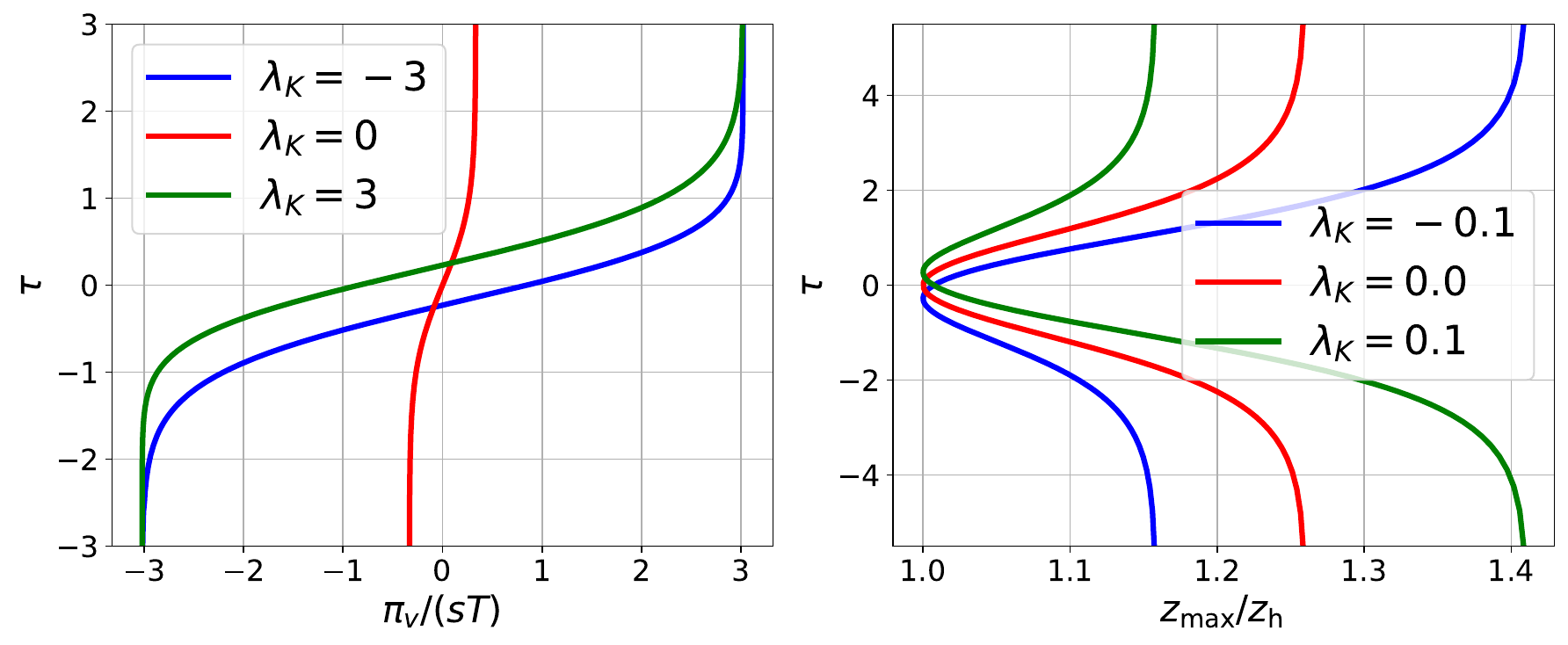}
    \caption{Anchoring time $\tau$ as a function of the conserved momentum $\pi_v$ (left) and the turning point $z_\mathrm{max}$ (right) for the $K$-observable at various values of 
    $\lambda_K$. Unlike the $C^2$-observable, $\tau$ is no longer antisymmetric in $\pi_v$ when $\lambda_K \neq 0$, and $\tau=0$ does not coincide with $\pi_v=0$. Furthermore, the 
    symmetry of $z_\mathrm{max}$ with respect to $\tau$ is broken: for $\lambda_K>0$, the surfaces bend towards the singularity of the white hole in region IV at early times, 
    while the opposite occurs for $\lambda_K<0$ (see Figure~\ref{fig:extremal_surfaces_K}). To highlight these effects, different sets of $\lambda_K$ values are used in the two plots.}
    \label{fig:tau_extrinsic}
\end{figure}
By numerically computing the integral in Equation~\eqref{eq:tau_momentum_relation_extrinsic} for different values of $\pi_v$, we are able to obtain a relation between $\tau$ and $\pi_v$. 
For the negative branch, the results are shown in the left panel of Figure~\ref{fig:tau_extrinsic}. Just as for the volume case $\lambda=0$, the relation is one-to-one. Since the potential
has no roots at a finite distance to the singularity if $\pi_v$ lies outside the critical range, the surfaces have no smooth turning point in this regime and can therefore not remain 
extremal when connecting the two boundaries, implying that no extremal surfaces exist in this regime.
The right panel shows $\tau$ as a function of the turning point $z_\text{max}$, which can be obtained from $u_+$. As can be seen, for $\lambda_K > 0$, the surface is able to approach the singularity of 
the white hole closely at early times, while this is not the case for the black hole singularity at late times. The opposite behavior occurs for $\lambda_K < 0$. 
We interpret this as follows: extremal surfaces obtained with the negative branch are at all times bent toward the past singularity for $\lambda_K>0$ and toward the future singularity for $\lambda_K<0$.
As a result, these surfaces provide deeper probes of the black hole or white-hole singularities, respectively. 

\begin{figure}[htbp!]
    \centering
    \includegraphics[width=0.65\textwidth]{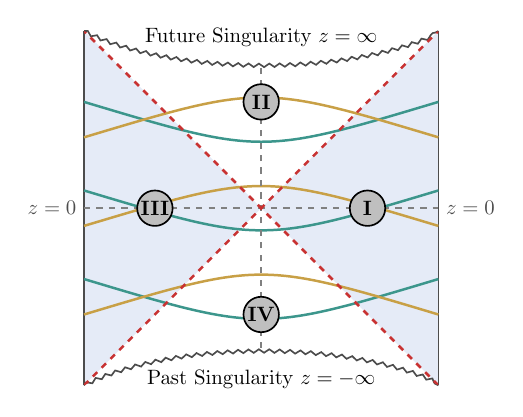}
 \caption{
    Extremal surfaces of the $K$-functional at different anchoring times. 
    Surfaces with $\lambda_K > 0$ (green) bend towards the past singularity across the diagram, avoiding the future singularity, 
    while surfaces with $\lambda_K < 0$ (brown) exhibit the opposite behavior, bending towards the future singularity. Nevertheless, the growth rate of the $K$-observable is the same for $\lambda_K$ and $-\lambda_K$ in the late and early time limit,
    as shown in Figure~\ref{fig:tau_extrinsic}.}
    \label{fig:extremal_surfaces_K}
\end{figure}
A schematic illustration of this behavior is shown in 
Figure~\ref{fig:extremal_surfaces_K}. The time evolution of extremal surfaces associated with the $K$-functional is qualitatively different from that of the $C^2$-functional. In the latter case, the extremal surfaces are reflection symmetric about the horizontal $t=0$ line, indicated by the dashed grey line in Figure~\ref{fig:extremal_surfaces_K}. This symmetry is characteristic of observables that do not depend on the embedding of the surface. The origin of this difference is that, for such observables, the potential $\mathcal{U}$ in Equation~\eqref{eq:eom_embedding} depends only on $\pi_v^2$, and the $\tau$-integral in Equation~\eqref{eq:tau_momentum_relation} is antisymmetric in $\pi_v$, which enforces reflection symmetry of the extremal surfaces.

\bibliographystyle{JHEP}
\bibliography{refs}

\end{document}